\documentclass{IOS-Book-Article}

\usepackage[T1]{fontenc}
\usepackage{titlesec}
\titleformat{\paragraph}[runin]
{\bfseries}{\theparagraph}{1em}{}

\usepackage{cite}
\usepackage{comment}
\usepackage{mathptmx}
\usepackage{amsmath}
\usepackage{adjustbox}
\usepackage{graphicx} % For including images
\usepackage{caption} % For customizing captions (optional)
\usepackage{subcaption} % For subfigures and captions
\usepackage{float}
\usepackage{booktabs}
\usepackage{fontawesome5}
\usepackage{pifont} % For checkmarks and crosses
\usepackage{xcolor}
\usepackage{multirow}
\usepackage{amsfonts}
\usepackage{appendix}
\usepackage[normalem]{ulem}
\usepackage{soul}\setuldepth{article}
\def\hb{\hbox to 11.5 cm{}}

\begin{document}

\pagestyle{headings}
\def\thepage{}
\begin{frontmatter}              % The preamble begins here.

%\pretitle{Pretitle}
\title{Hybrid Retrieval for Hallucination Mitigation in Large Language Models: A Comparative Analysis}

% Systematic Analysis of Retriever Impact on Hallucinations Generated by Large Language Models
% "Optimizing Retrieval-Augmented Generation: Reducing Hallucinations in Large Language Models with Hybrid Retrieval"
% "Hybrid Retrieval for Hallucination Reduction in Large Language Models: A Systematic Evaluation"
% "Enhancing Context Retrieval to Mitigate Hallucinations in LLMs: A Comparative Study"
% "Reducing Hallucinations in Retrieval-Augmented Generation Systems via Hybrid Search Optimization"
% "The Role of Hybrid Retrieval in Improving Answer Accuracy and Reducing Hallucinations in LLMs"

\markboth{}{January 2025\hb}

\author[A, B]{\fnms{Chandana sree} \snm{Mala}%
\thanks{Corresponding Author: Chandana sree Mala, email: [c.mala@studenti.unipi.it, chandana.mala@sns.it], ORCiD: https://orcid.org/0009-0004-7500-6121}}
\author[B]{\fnms{Gizem} \snm{Gezici}}%
\author[B]{\fnms{  Fosca} \snm{Giannotti}}

\runningauthor{C. Mala et al.}
\address[A]{Department of Computer Science, University of Pisa}
\address[B]{Department of Computer Science, Scuola Normale Superiore}

\begin{abstract}

Large Language Models (LLMs) excel in language comprehension and generation but are prone to hallucinations, producing factually incorrect or unsupported outputs. Retrieval-Augmented Generation (RAG) systems mitigate this by grounding LLM responses with external knowledge. This study evaluates the relationship between retriever effectiveness and hallucination reduction in LLMs using three retrieval approaches: sparse retrieval (BM25-based keyword search), dense retrieval (semantic search with Sentence Transformers), and the proposed hybrid retrieval module which incorporates information from query expansion and further fuses the results of sparse and dense retrievers through a dynamically-weighted Reciprocal Rank Fusion (RRF) score.
Using the HaluBench dataset, a benchmark for hallucinations in Question Answering tasks, we assess retrieval performance with MAP and NDCG metrics, focusing on the relevance of the top-3 retrieved documents. Results show that the hybrid retriever has a better relevance score outperforming both sparse and dense retrievers. Further evaluation of LLM-generated answers against ground truth using metrics like accuracy, hallucination rate, and rejection rate reveals that the hybrid retriever achieves the highest accuracy on fails, the lowest hallucination rate, and the lowest rejection rate. These findings highlight the hybrid retriever's ability to enhance retrieval relevance, reduce hallucination rates, and improve LLM reliability, emphasizing the importance of advanced retrieval techniques in mitigating hallucinations and improving response accuracy.

\end{abstract}

\end{frontmatter}
\markboth{January 2025\hb}{January 2025\hb}

\section{Introduction}
\label{sec:intro}

Advancements in natural language processing (NLP) have brought large language models to the forefront, revolutionizing both academic research and practical applications in diverse domains.
RAG is an approach that enhances LLMs by integrating retrieval mechanisms to improve response accuracy and reduce hallucinations~\cite{lewis2020retrieval}. 
%Recently, retrieval augmented generation , as a typical and representative LLM-enabled system, has emerged to further enhance the capabilities of LLMs.
%
Instead of relying solely on the model’s internal knowledge, RAG retrieves relevant external documents from a knowledge source (e.g., databases, search engines, or vector stores) and incorporates them into the generation process.
By integrating retrieval mechanisms from external sources, RAG effectively addresses major limitations of standalone LLMs~\cite{zhao2024towards, chen2024benchmarking}, including the high costs associated with training and fine-tuning~\cite{touvron2023llama}, the issue of hallucination~\cite{huang2023survey, zhang2024siren, bai2024hallucination, bechard2024reducing}, and constraints imposed by the input window~\cite{vaswani2017attention} and knowledge cut-off\cite{lewis2020retrieval}.
%In this way, RAG effectively mitigates key limitations of standalone LLMs~\cite{zhao2024towards, chen2024benchmarking}, such as the high cost of training and fine-tuning \cite{touvron2023llama}, the phenomenon of hallucination \cite{huang2023survey, zhang2024siren, bai2024hallucination} and the limitations inherent in the input window \cite{vaswani2017attention}. 
%By retrieving relevant and up-to-date information from external knowledge bases, Given the promising performance and cost effectiveness, RAG has achieved success across various tasks and scenarios, including chatbots, factual QA [8], longform generation \cite{chen2023understanding}, and automatic code completion [10][13].
Moreover, RAG has already become a foundational technology in various real-world products like Contextual AI [14] and Cohere [15].
%
%RAG is a method of text generation that draws from both dynamically retrieved external knowledge and patterns learned during training \cite{merth2024superposition}.
%\textcolor{red}{I did not like this definition to be honest - we cannot say that it is text generation it is rather a retrieval system}

RAG system blends the encyclopedic memory of a search engine with the generative models and consists of two main modules as the retrieval phase (R) and the generation phase (G). In the retrieval phase, a retriever fetches relevant documents based on the input query using three retrieval approaches: a sparse retriever leveraging ($BM25$~\cite{robertson2009probabilistic}-based lexical matching), a dense retriever(using embeddings from Sentence Transformers), or a hybrid approach (combining both methods). These retrieval algorithms have been inspired from Information Retrieval (IR), where search systems seek for alternative retrieval approaches to satisfy the information need of users, i.e. retrieving the most relevant documents at the top positions of a ranked list with respect to a given user query\cite{sawarkar2024blendedragimprovingrag}. 
Many popular web search engines employ $BM25$ or similar ranking algorithms to determine the relevance of search results for a given query.
%
%In the generation phase, the retrieved documents are provided as context to the chosen LLM, which generates a response using both its internal knowledge and the retrieved information.
%which stands for the size and type of LLM, and the Retriever (R) are the two essential elements that make up the RAG systems.
%These two main modules of RAG, the retrieval and the generation work harmoniously together since the retriever and the LLM have complementary capabilities.
%While the LLM may not have access to all relevant data, it can effortlessly generate sentences. This is where the retriever plays a crucial role, efficiently scanning vast collections of internal documents (external knowledge base) to find relevant information that enhances and informs the model’s output.
%Without the retriever, RAG would be like a well-spoken individual who delivers irrelevant information~\cite{sawarkar2024blended}. Search has been a primary focus of research in information retrieval, with numerous studies investigating alternative approaches. Historically, the BM25 (Best Match) algorithm, which employs similarity search, has been a cornerstone in this subject . BM25 prioritises pages based on their relevance to a query, using Term Frequency (TF), Inverse Document Frequency (IDF), and Document Length to calculate a relevance score.  

This paper explores the effectiveness of different retrieval methods in reducing hallucinations.
Note that~\emph{hallucinations} occur when the generated answers are not faithful to the context (intrinsic hallucinations) or don’t align with factual reality (extrinsic hallucinations)~\cite{jian-etal-2022-embedding , ji2023survey}. In this paper, we focus solely on intrinsic hallucinations since in real-world settings, user-provided documents may contain information that conflicts with external knowledge sources.

To the best of our knowledge, this is the first study that evaluates the hybrid retrieval performance in mitigating hallucinations.
%\textcolor{red}{GIZEM: I could not understand the explanation here! Why we are only focusing on intrinsic hallucinations}.
%\textcolor{green}{Chandana: The terms intrinsic hallucination and extrinsic hallucination are used to categorize different types of errors or inaccuracies in the outputs of LLMs. 1. Intrinsic Hallucination: They occurs when the model generates information that contradicts or is inconsistent with the provided input or context. 2. Extrinsic Hallucination: They occurs when the model generates information that cannot be verified or is not supported by the provided input or context, but does not directly contradict it. So, in our paper we only focus on Intrinsic Hallucinations only}.
%
Our main contributions are as follows:

\begin{itemize}
\item We use a query expansion module to increase the coverage of the hybrid retrieval phase.

\item We evaluated how different types of retrieval performance affect hallucinations in LLM generated outputs.
\end{itemize}

This paper is organized by introducing the motivation behind reducing hallucinations in LLMs through Retrieval-Augmented Generation. The second section surveys recent RAG studies, highlighting key retrieval strategies and their relevance to mitigating hallucinations. In the third section, we detail our hybrid retrieval methodology, underscoring query expansion and dynamic weighting. The fourth section outlines the experimental setup and results on the dataset, and the paper concludes with final observations on the effectiveness of the proposed hybrid retriever followed by future work.

\section{Related Work}

RAG systems have emerged as a promising solution to the inherent limitations of LLMs, particularly their tendency to hallucinate or generate inaccurate information \cite{semnani-etal-2023-wikichat, chang-etal-2024-detecting}. By integrating retrieval mechanisms, RAG systems retrieve relevant external knowledge during the retrieval phase, which is then incorporated into the query. This ensures that the LLM's generated output is informed by up-to-date and contextually relevant information\cite{wang2024searching}. 

Early work in \cite{shuster2021retrieval} and \cite{bechard2024reducing} demonstrated that complementing LLMs with specialized retrievers can substantially ground the generated text in factual evidence.This has spurred research into a variety of domain-specific and application-specific RAG approaches, such as \cite{shi2024ask, anjum2024halo}, where sophisticated modules decrease hallucinations by parsing industry abbreviations and consolidating context from heterogeneous sources.

Additionally, \cite{salemi2024evaluating, Gao2023RetrievalAugmentedGF} and  \cite{li2025enhancing, pmlr-v119-guu20a} illustrate both benchmark comparisons and methodological guides for improving retrieval accuracy, with an emphasis on ensuring that even black-box LLMs can trace back to reliable evidence like discussed in this paper\cite{shi-etal-2024-replug}.

Recent research has focused on enhancing the efficiency and performance of RAG systems by improving their retrieval components like discussed in this papers\cite{sawarkar2024blended} and \cite{arivazhagan2023hybrid, wang2024searching} highlight how fusing dense and sparse retrieval signals yields higher relevance in challenging Q\&A contexts\cite{omrani2024hybrid}.

This fusion approach is further explored in \cite{bruch2023analysis} and \cite{rackauckas2024rag, kalra2024hypa}, where rank fusion, weighted scoring, and dynamic weighting strategies emerge as key factors for precise, context-rich retrieval. Contributions such as \cite{zhao2024towards, chen2024benchmarking, hsia2024ragged, olufade2021dynamic} and offer an analytical lens through which prompt optimization, domain adaptation, and query expansion recommender modules, most recent paper\cite{li2025enhancing} demonstrate that by expanding the query to relevant fields may enhance response quality by improving the relevance of the retrieved information which can further reduce irrelevance or hallucinations. 

Despite considerable progress in hybrid retrieval and RAG systems, gaps remain in understanding how retrieval approaches dynamically adapt to specific query scenarios and how these adaptations influence hallucination reduction. By extending the findings of previous research, our study systematically investigates the role of hybrid retrieval in mitigating hallucinations, ultimately paving the way for more reliable and accurate outputs in large language models.

\section{Methodology}
\label{sec:method}
In this section we describe our RAG system which is composed of two main modules as the retrieval and the generation phase as mentioned in Section~\ref{sec:intro}. In the retrieval phase, differently from the studies in the literature, we incorporated a query expansion ($QE$) module on top of the hybrid retrieval. The goal of this step is to address~\emph{lexical chasm}, i.e. the gap or the mismatch between the vocabulary used to formulate query and to represent information in documents.

%Our methodology unfolds in a sequence of three progressive steps: Dataset Preparation, Retrieval Phase and Generation Phase. 
%In a RAG pipeline\cite{ravi2024lynx}, we first (i) Retrieve the relevant document given a query, then (ii) Generate an answer to the query given the retrieved chunks and query passing them as context(s) to an LLM

% \begin{figure}
%     \centering
%     \includegraphics[width=\linewidth]{images/final_pipeline.pdf}
%     \caption{Caption}
%     \label{fig:RAG_pipeline}
% \end{figure}

\begin{figure}
    \centering
    \includegraphics[trim=0 180 0 120,clip,width=\linewidth]
    {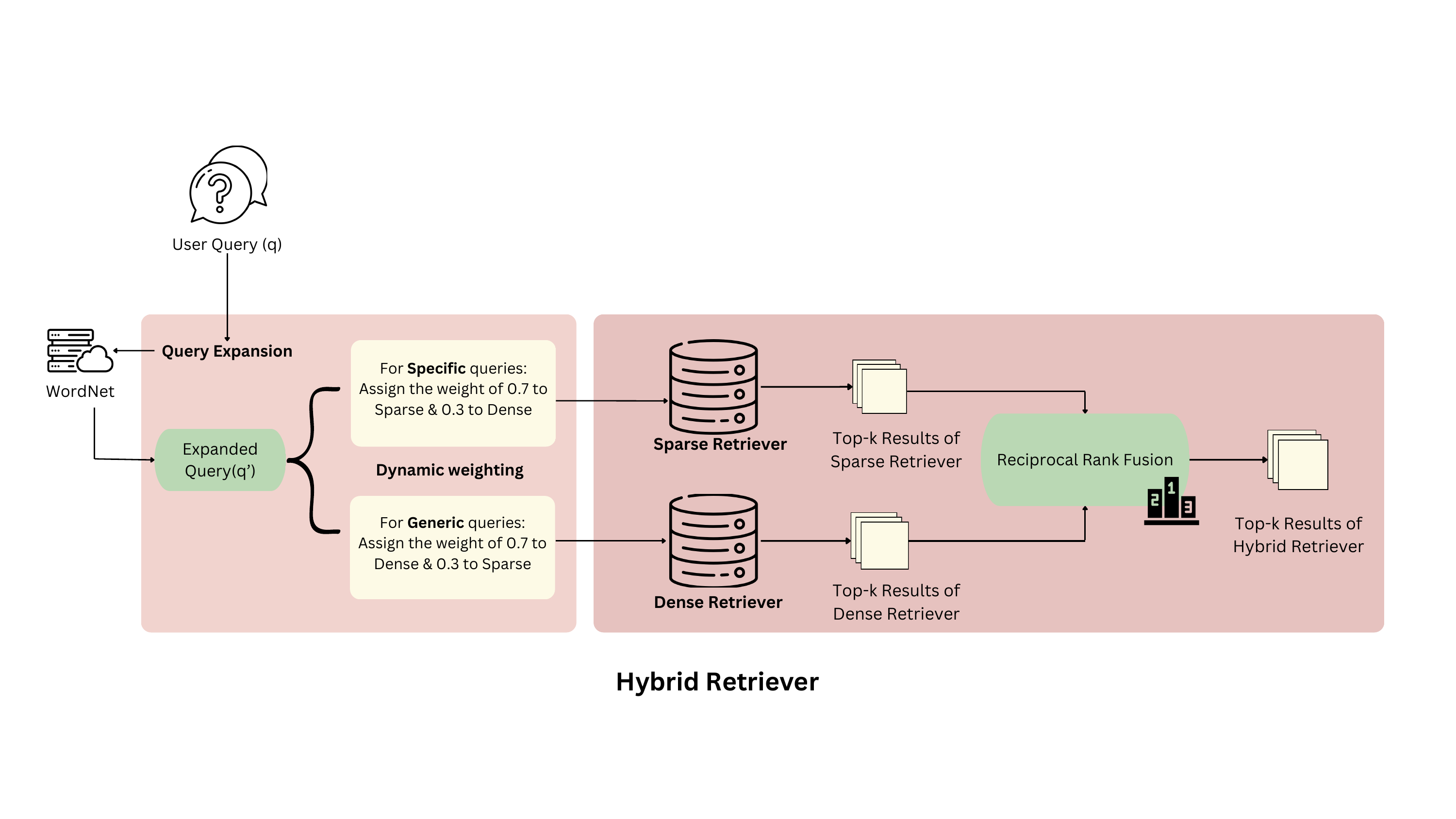}
    \caption{Our Hybrid Retriever Pipeline}
    \label{fig:hybrid_pipeline}
\end{figure}

\begin{figure}
    \centering
    \includegraphics[trim=0 50 0 0,clip,width=0.8\linewidth]{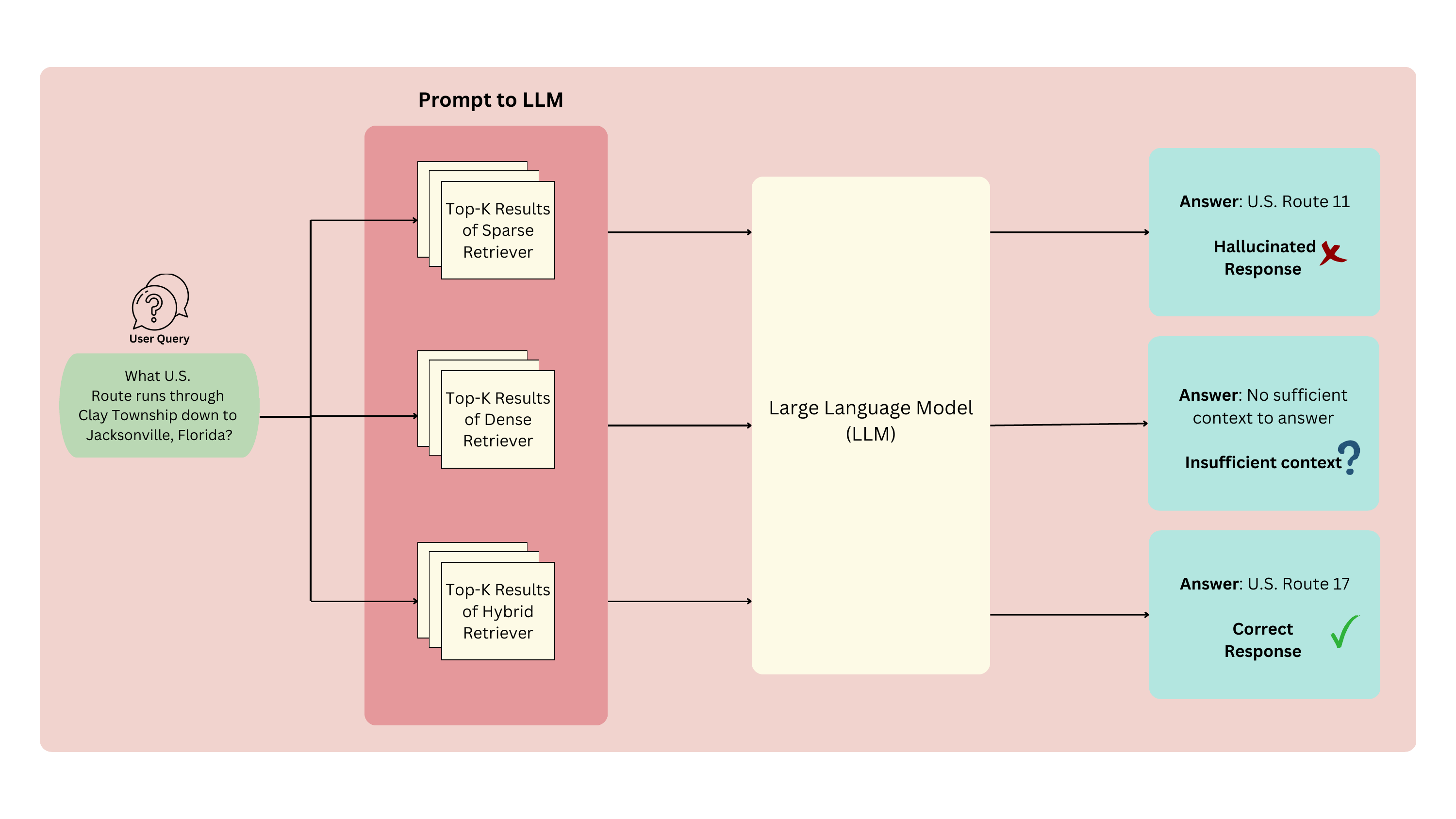}
    \vspace{-1em}
    \caption{Generation phase}
    \label{fig:generation_pipeline}

\end{figure}

\subsection{\textbf{Retrieval Phase}}
\label{subsec:retrieval}

The retrieval phase of a LLM-driven RAG system often contains two main components: the~\emph{indexed database} and the~\emph{retriever}~\cite{zhao2024towards}. The~\emph{indexed database} $DB$ is an external knowledge-base which is a structured collection of documents $d_i \in D$, for $i=\{1,...,n\}$. These documents include domain-specific knowledge, thus the relevant information with respect to the potential user queries of the current use-case.
%Organized and optimized for efficient searching, it serves as the foundation for retrieving pertinent information.
%
The steps in the retrieval phase are as follows. First, $D$ is stored offline in $DB$. Then, the~\emph{retriever} encodes $q$ and all $D$ in a vector space. 
Following that the~\emph{retriever} applies a chosen similarity function $f_{sim}$ which computes a similarity score between two given vector representations of $q$ and $d_i$ and ranks $d_i$ based on their relevance to a given query $q$. In this way, the most relevant $d_i$ with respect to $q$ is supposed to get the highest score.
 
%Given a query \emph{Q} and the \textit{indexed database} comprising \emph{n} documents \emph{D1, D2, . . . , Dn}, the \emph{retriever} identifies the order of these documents according to a similarity criterion \emph{sim}. Formally, 

\begin{comment}
\begin{equation}
D_{i1}, D_{i2}, \ldots, D_{in} = \text{sort}(D_1, D_2, \ldots, D_n, \text{sim}(Q, D_i))
\end{equation}
\end{comment}

Various retrieval methods leverage different types of information from $q$ and $D$. Sparse retriever ($Ret_S$) performs a~\emph{keyword search} through projecting $q$ and $D$ into a sparse vector space, usually employing traditional Bag-of-Words (BoW) techniques like $BM25$~\cite{robertson2009probabilistic} or $tf*idf$. These BoW approches often struggle with synonyms and 
varying contextual meanings and fails to capture the semantic relationships between the words.

To address these limitations, dense retrievers ($Ret_D$)~\cite{izacard2021unsupervised, karpukhin2020dense} perform~\emph{semantic search} by encoding $q$ and $D$ into dense vectors to capture their semantic meaning.
%
%based on their utilization of sparse or dense, or both combined representations of queries and documents \cite{lin2021pyserini}.
%To address the limitations of these methods, which , dense retrievers calculate similarity scores by encoding queries and documents into dense vectors that capture their semantic meaning.
%In Sparse retriever, keyword search is performed by computing similarity by projecting the query and documents into a sparse vector space that aligns with the vocabulary of the documents, typically using traditional Bag-of-Words methods such as TF-IDF or BM25 \cite{robertson2009probabilistic}. To overcome the limitations of these methods, which can struggle with synonyms or varying contextual meanings, dense retrievers \cite{izacard2021unsupervised , karpukhin2020dense},  obtain similarity scores by encoding queries and documents as dense vectors that capture their semantic meaning and generate dense embeddings for tasks such as semantic search, information retrieval, and clustering.

%
On the other hand, hybrid approach leverages information both from sparse and dense vector representations through combining their similarity (relevance) scores.
While the conventional approach for hybrid retriever typically uses a linear combination of sparse and dense retriever scores, our hybrid retriever denoted as $Ret_{Hyb-{RRF}}$ utilizes Reciprocal Rank Fusion (RRF)~\cite{cormack2009reciprocal, bruch2023analysis} to establish the final ranking.
In contrast to score-based interpolation, RRF uses the ranking positions of each document retrieved by the individual retrievers, providing a more balanced and effective fusion of results.
Furthermore, rather than choosing a retriever among
sparse, dense, or hybrid retrieval strategies, our proposed retriever $Ret_{Hyb-{RRF}}$ compares all these three strategies and adapts its behaviour based on the current query's characteristics.
Unlike many hybrid models that rely on computationally intensive dense retrievers requiring complex compression techniques such as linear projection, PCA, or product quantization~\cite{luo2022study}, $Ret_{Hyb-{RRF}}$ enhances retrieval effectiveness by integrating a query expansion ($QE$) module to increase the query coverage and adapting the weights of different retrieval approaches with respect to the query's characteristics.
%\textcolor{red}{GIZEM: Okay but after I read the next section we dont need to say WordNet here it is a detail lets leave this detail to the next section.} \textcolor{green}{Chandana: okay, done}
%\textcolor{green}{Chandana:In general, in traditional Hybrid retrieval process they just take the results of multiple retrievers and re ranks them to give a better ranked results, but in our methodology/approach we are enhancing the hybrid retrieval process by incorporated query expansion and weighting techniques so our modified hybrid retrieval works even better than the traditional one for retrieving the appropriate results(for example:- " https://medium.com/@shubhamsarkar996/hybrid-search-in-rag-concept-of-weighted-reciprocal-rank-fusion-rrf-part-1-ae570d9c1879 " see this article, even when the traditional hybrid retrieval fails in some cases our weighting approach might work better. this is what I wanted to convey in this paragrapgh here.}

In this work, our aim is to systematically evaluate three retrieval approaches ~\emph{sparse},~\emph{dense}, and ~\emph{hybrid} to measure their effectiveness in mitigating hallucinations. The hybrid method integrates keyword and semantic searches through query expansion and dynamic weighting as illustrated in Figure \ref{fig:hybrid_pipeline}, aiming to maximize both precision and recall~\cite{lin2021pyserini} and further examine its influence on LLM generated responses as illustrated in Figure \ref{fig:generation_pipeline}

\paragraph{Hybrid retrieval approach}
Our hybrid retrieval process $Ret_{Hyb-{RRF}}$ starts with $QE$, an essential step aimed at enhancing the retrieval phase by augmenting $q$ with semantically related terms. For this purpose, WordNet~\cite{miller1995wordnet}, a comprehensive lexical database that demonstrates the relationships between words—such as synonyms (similar meanings), antonyms (opposite meanings), or words within the same category—is utilized. 
%
%Lexical chasm between queries and document titles is the main obstacle for improving base relevance of a search engine. This is substantially rooted from two things: i) authors of queries and documents are di erent (e.g. diverse vocabulary usage), ii) insu cient knowledge of technical terms in the corresponding domain.

%The hybrid retrieval process begins with query expansion, a critical step designed to improve the retrieval of relevant documents by enriching the original query with semantically related terms. This is achieved using WordNet\cite{miller1995wordnet}, a lexical database that provides synonyms, and other semantic relationships for words. The goal of this step is to address the vocabulary mismatch problem, where the query and relevant documents may use different but semantically related terms.
%
Let the original query $q$ be seen as the set of query terms $q_j$, denoted as $q_j \in q$, for $j=\{1,...,|q|\}$ where $|q|$ is the number of terms in the query. In $QE$, for each $q_j$, we retrieve a set of synonym terms from WordNet via NLTK\footnote{\url{https://www.nltk.org/}} and use only $top-2$ most-relevant terms denoted as $T(q_j)$ to expand $q$ not to change its original intent.
Then, the expanded query $q'$ is defined as:

\begin{equation}
q' = q \cup T(q_j) 
\end{equation} 

\begin{comment}
\begin{equation}
q' = \bigcup_{i=1}^{n} S(q_i)   
\end{equation} 
\end{comment}

%where $S(q_i)={q_i} \cup Synonyms(q_i)$ 

%
As an example, if $q_j = car$, we can include $T = \{automobile, vehicle\}$ from WordNet, to create $q'$. Then $q'$ is utilized during $Ret_{Hyb-{RRF}}$ to close the lexical gap between $q$ and $d_i$. Query expansion techniques have already been shown to enhance recall in information retrieval tasks~\cite{carpineto2012} through increasing query coverage.
%ensure that documents with synonyms and semantically related terms are included. 
%For example, if the query term is "car," WordNet might expand it to include synonyms such as "automobile" and "vehicle." This expanded query is then used in hybrid retrieval stage to ensure that documents containing synonymous or related terms are not missed. Query expansion has been shown to improve recall in information retrieval tasks, as discussed in \cite{10.1145/2071389.2071390}
%
After the $QE$, $Ret_{Hyb-{RRF}}$ employs dynamic weighting~\cite{azad2019query} to optimize the contributions of $Ret_S$ and $Ret_D$ based on the characteristics of $q'$. These characteristics are assessed by evaluating the term distribution and level of informativeness of $q'$~\cite{SprckJones2021ASI}.
\emph{Specific} queries that are detailed, focused, and often seek precise information or exact matches are given greater weight to $Ret_S$, whereas \emph{general} queries that are broad or open-ended which lack specific details and typically require a high-level or conceptual information, are weighted more to $Ret_D$~\cite{kuzi2020leveraging}.
%The specificity of a query is determined by analyzing its term distribution and informativeness.
%Specific queries, which are more likely to require exact matches, are assigned higher weights to the sparse retriever, while general queries, which benefit from conceptual understanding, are assigned higher weights to the dense retriever\cite{kuzi2020leveraging}.

Let $w_{Ret_S}$ and $w_{Ret_D}$ represent the weights assigned to $Ret_S$ and $Ret_D$. These weights are dynamically computed by $Ret_{Hyb-{RRF}}$ based on a query specificity score~\cite{salton1988term, aizawa2003information, SprckJones2021ASI} denoted as $S(q')$: 

\begin{equation}
\label{eq:specifity}
    S(q') = \frac{1}{|q'|} \sum_{i=1}^{|q'|} tf*idf(q_j)
\end{equation}

Then, we assign the weights to retrievers $w_{Ret}$ based on the query specificity score as follows:

\begin{align}
w_{Ret_S} = \alpha S(q') \\
 w_{Ret_D}  = 1 - w_{Ret_S}
\end{align}

\begin{comment}
The weights are then assigned as follows:

$
w_{\text{BM25}} = \alpha \cdot S(Q), \quad w_{\text{Semantic}} = 1 - w_{\text{BM25}}
$
\end{comment}

%\textcolor{red}{GIZEM: What is alpha value we chose?}
%\textcolor{green}{Chandana: Its just a scaling factor used for normalization, The default value for aplha is typically 1, assuming S(Q') is normalized to [0,1]}

where $\alpha$ that is set to 1 by default, serves as a scaling factor for normalization.
%to ensure that the weights add up to 1. 
%
For specific queries with a high specificity score $S(q')$, $w_{Ret_S}$ will be higher, whereas for general queries with a low $S(q')$, $w_{Ret_D}$ will be lower.
This dynamic weighting mechanism customizes the retrieval process based on the query's characteristics, potentially enhancing both precision and recall.
%For specific queries (high $S(Q)$), $w_{\text{BM25}}$ is higher, while for general queries (low $S(Q)$), $w_{\text{Semantic}}$ is higher. This dynamic weighting mechanism ensures that the retrieval process is tailored to the nature of the query, improving both precision and recall.

%\textcolor{blue}{GIZEM: LEFT HERE!!!}

%
Next, ${Ret_S}$ and ${Ret_D}$ independently retrieve the $top-k$ ($k=3$, in our case) documents denoted as $D_{Ret_S}$ and $D_{Ret_D}$ based on their respective scoring mechanisms, $BM25$ for ${Ret_S}$ using exact lexical matches and $cosine-similarity$ for ${Ret_D}$ which aims to capture semantic similarity. For ${Ret_D}$, the vector embeddings of $q'$ and $D$ are both dense representations created by the model ~\emph{sentence-transformers/all-mpnet-base-v2}\footnote{\url{https://huggingface.co/sentence-transformers/all-mpnet-base-v2}} \cite{reimers2019sentence}.
Note that $BM25$ is particularly effective for specific queries, while $cosine-similarity$ is more effective for general queries.
More details and the mathematical formula of $BM25$ can be found in~\cite{manning2009introduction} and a detailed discussion on the use of sentence embeddings for semantic search in~\cite{karpukhin2020dense}.

After retrieving $D_{Ret_S}$ and $D_{Ret_D}$, these two ranked lists are fused using a weighted RRF score denoted as $RRF_{weighted}$ which is computed as follows:

\begin{equation}
RRF_{weighted}(d_i) = \sum_{Ret \in \{Ret_S, Ret_D\}} \frac{w_{Ret}}{\epsilon + r_{Ret}(d_i)}
\end{equation}

where $r_{Ret}(d_i)$ is the rank of $d_i$ which exists in $D_{Ret_S}$ or $D_{Ret_D}$, and $w_{Ret}$ is the weight assigned to the respective retriever during the dynamic weighting step of $Ret_{Hyb-{RRF}}$, and $\epsilon$ is a small constant to avoid division by zero. $RRF_{weighted}(d_i)$ is the relevance score then utilized by $Ret_{Hyb-{RRF}}$ to rank the documents $d_i$ with respect to a given query $q$.
\begin{comment}
\begin{itemize}
    \item $\text{rank}_r(D)$ 
    \item $w_r$ is the weight assigned to the retriever (computed in the dynamic weighting step),
    \item $k$ is a small constant to prevent division by zero.
\end{itemize}
\end{comment}
%
%Based on these,  higher scores The RRF method is described in \cite{cormack2009reciprocal, Bruch2013}. \textcolor{red}{GIZEM: If these two works show this weighted function, then cite them here otherwise cite them above when we first define RRF. I could not understand this part: while also accounting for the dynamic weights assigned earlier.}
%\textcolor{green}{Chandana: okay,noted}
%
As the final step, $top-k$ documents with the highest $RRF_{weighted}(d_i)$ denoted as $D_{Ret_{Hyb-{RRF}}}$ are obtained by $Ret_{Hyb-{RRF}}$ as follows:

\begin{equation}
D_{Ret_{Hyb-{RRF}}} = \arg\max_{d_i} RRF_{weighted}(d_i)
\end{equation}

Thus, $d_i$ which are highly ranked by both $Ret_S$ and $Ret_D$ will receive higher relevance scores by $Ret_{Hyb-{RRF}}$, while incorporating the previously assigned dynamic weights $w_{Ret_S}$ and $w_{Ret_D}$.
$D_{Ret_{Hyb-{RRF}}}$ is expected to provide the most relevant (precision) and the broadest context (recall) for $q$, leveraging the advantages of both lexical ($Ret_S$) and semantic retrieval ($Ret_D$) methods, where $|D_{Ret_{Hyb-{RRF}}}| = 3$ (number of the retrieved documents by the hybrid retriever).
This step ensures that the retrieval process not only identifies relevant documents but also ranks them in a way that maximizes their utility for downstream tasks, such as answer generation in RAG systems.

\subsection{\textbf{Generation Phase}}
\label{subsec:generation}

The generation phase consists of two key components: a prompt $p$ and a chosen pre-trained LLM $M$. In the retrieval phase, $Ret_{Hyb-{RRF}}$ retrieves the $top-3$ most-relevant documents $D_{Ret_{Hyb-{RRF}}}$ from $DB$ based on the query $q'$ (expanded version of $q$) to incorporate context information into the query for the generation phase. For this, $q'$ is concatenated with $D_{Ret_{Hyb-{RRF}}}$ to form $p$ (see Appendix~\ref{sec:prompt-details} for the prompt we use) which is then used as a prompt for $M$ to generate a response to the original query $Q$.
We use standard prompting with detailed instructions in zero-shot settings, i.e. without providing any exemplars, although alternative approaches including few-shot learning~\cite{brown2020language} or Chain of Thought (CoT) prompting~\cite{wei2022chain}, i.e. step-by-step reasoning, can be used in RAG systems as demonstrated by~\cite{wang2024rat}.

Note that we use the pre-trained LLM without any fine-tuning, i.e changing the model weights. The model used for the response generation is \texttt{LLaMA-3-8B-Instruct}\cite{touvron2023llama}, is a cutting-edge large language model with 8 billion parameters, \texttt{max new tokens = 8132, temperature=0.8, top p=0.9} optimized for instruction-following tasks.

\section{Experimental Setup}
In this section, we outline the experimental setup based on the proposed RAG pipeline, as detailed in Section~\ref{sec:method}. To evaluate if the proposed pipeline is a promising approach on mitigating hallucinations, we separately assess retrieval performance (Section~\ref{subsec:retrieval}) and the overall effectiveness of the RAG pipeline by examining both the retrieval phase output and the final response to the query $q$, which integrates the results of the retrieval and generation phases (Section~\ref{subsec:generation}). This approach enables us to assess how retrieval performance impacts the overall effectiveness of the pipeline in mitigating hallucinations.

\subsection{Dataset}
\label{subsec:dataset}
We conduct our study on the HaluBench dataset~\cite{ravi2024lynx}, a comprehensive hallucination evaluation benchmark consisting of 13,867 samples. The dataset is a combination of six diverse benchmarks that are source datasets, i.e. DROP~\cite{dua2019drop}, HaluEval~\cite{li2023halueval}, RAGTruth~\cite{niu2023ragtruth}, FinanceBench~\cite{islam2023financebench}, PubMedQA~\cite{jin2019pubmedqa}, and COVIDQA~\cite{moller2020covid}, and contains hallucinated and faithful responses to questions that may span various domains, including general knowledge, reasoning, specific facts, or specialized topics including finance and healthcare. The HaluBench dataset includes examples of challenging-to-detect hallucinations, meaning instances that seem plausible but are not faithful to the context.
Each data instance in HaluBench includes a context passage ($d_i$), a question based on that context ($q_{d_i}$), an LLM-generated answer, and a binary label indicating whether the answer constitutes a hallucination in relation to the context (\texttt{PASS} for correct answers and \texttt{FAIL} for hallucinated answers). The binary labels in HaluBench were generated by comparing the LLM-generated answer with the ground truth from the source dataset. Therefore, for our evaluations, if an instance is labeled as \texttt{PASS}, its LLM-generated answer was considered the ground truth. However, for instances labeled as \texttt{FAIL}, the ground truth answer was directly obtained from the corresponding source dataset.

For the evaluations, we utilized the different versions of HaluBench to separately assess the retrieval phase and the overall effectiveness of the RAG pipeline in reducing hallucinations. To evaluate the retrieval performance, we employed the entire HaluBench dataset, which contains 13,867 instances denoted as $HaluBench_{orig}$. In this dataset, we measured the performance of the retrievers in an automated manner by using the questions $q$, and the respective context passages $q_{d_i}$. If a given retriever retrieves the context passages $q_{d_i}$ for a given $q$, then these retrieved documents are $relevant$, otherwise $irrelevant$.
On the other hand, for assessing the overall performance in mitigating hallucinations, the evaluation cannot be fulfilled in an automated manner since the evaluation requires reasoning capabilities and should be done by a human annotator which is the standard approach~\cite{ravi2024lynx}.
Thus, we could not annotate the entire dataset and used a randomly sampled subset of 300 instances which is denoted as $HaluBench_{small}$, with 50 instances from each of the six source datasets to maintain dataset diversity. This subset contained an equal number of~\texttt{PASS} and~\texttt{FAIL} instances per source dataset, with 25 instances of each, ensuring a balanced evaluation.

The responses of the entire RAG pipeline for all the queries $Q$ in the annotated dataset were labelled by a human annotator through comparing the generated response and the ground truth answer with the following three labels:

\begin{itemize}
    \item Hallucinated Answer (\textcolor{red}{\ding{55}}): The generated answer is factually incorrect or unsupported by the provided context.
    \item Correct Answer (\textcolor{green}{\ding{51}}): The generated answer matches the ground truth and is factually accurate.
    \item Insufficient Context (\textcolor{blue}{\textbf{?}}): The retrieved context does not provide sufficient information to answer the query.
\end{itemize}

For the comparative evaluation, the responses from all three RAG pipelines, which differ in their retrieval approaches (sparse, dense, and hybrid), were fully annotated. The annotated files can be found in our ~\emph{github repo}\footnote{\url{https://anonymous.4open.science/r/HybridRAG\_for\_Hallucinations-884F}}
Note that $HaluBench_{small}$ was annotated by a single human annotator due to time constraints. Nonetheless, the query set we annotated was not difficult so we believe that it was less prone to disagreements. 10 samples from the annotated dataset can be found in the Appendix.
\subsection{Evaluation Metrics}
\label{subsc:evalmetrics}

\paragraph{Retrieval Metrics}
\label{retrieval_metrics}

To evaluate the retrieval performance of our hybrid approach $Ret_{Hyb-{RRF}}$, we compared its performance with the sparse ${Ret_S}$, and dense ${Ret_D}$ retrievers. For this, we used commonly-used order-aware metrics from the Information Retrieval (IR) domain, namely Mean Average Precision (MAP)\cite{salemi2024evaluating,10.1145/1277741.1277790, yu2024evaluation} and Normalized Discounted Cumulative Gain (NDCG)\cite{salemi2024evaluating,10.1145/582415.582418, sawarkar2024blendedragimprovingrag}.

As mentioned in Section~\ref{sec:method}, since each retriever returns a ranked list of three documents ($top-3$), these metrics were computed at a cut-off value, $k=3$, i.e. number of documents considered for the evaluation.

MAP averages the $precision@k$ metric at each relevant item position in the retrieved ranked list of documents, where $precision@k$ measures the proportion of relevant documents in a ranked list of size $k$.
For a query $q$, the Average Precision (AP) is defined as:

\begin{equation}
AP = \frac{1}{|\text{Rel}_q|} \sum_{i=1}^{n} Precision@i \cdot \mathbb{1}[\text{rel}_i = 1]
\end{equation}

where $\mathbb{1}[\cdot]$ is the indicator function that specifies whether the document at rank $i$ is relevant and $|{Rel}_q|$ is the total number of relevant documents for query $q$. The MAP of a retriever is then computed as the mean of AP across the set of all queries $Q$ in the dataset as follows:

\begin{equation}
MAP = \frac{1}{|Q|} \sum_{q \in Q} AP(q)
\end{equation}

This metric rewards the retrieval approaches that put more relevant documents at the top of the ranked list.
DCG has a stronger concept of ranking which discounts the “value” of each relevant document based on its rank in a ranked list of size $k$ using a logarithmic discount function as follows:

\begin{align}
DCG@k = \sum_{i=1}^{k} \frac{2^{rel_i} - 1}{\log_2(i + 1)} \\
NDCG@k = \frac{DCG@k}{IDCG@k}
\end{align}

where $rel_i$ is the relevance grade of a document at rank $i$ and for the binary case, if a document is relevant, relevance grade is assigned as $1$, otherwise $0$.
NDCG@k then normalizes DCG@k by the “ideal” ranked list (IDCG@k), where every relevant document is ranked at the start of the list. For both MAP and NDCG metrics, higher scores mean better retrieval performance.

\paragraph{Overall Evaluation Metrics}
\label{overall_metrics}

To evaluate the overall performance of the RAG pipeline in mitigating hallucinations, we utilize the following metrics as defined in~\cite{yu2024evaluation, chen2024benchmarking}:

\begin{itemize}
    \item \textbf{Accuracy:\cite{li2023halueval, kalra2024hypa}} The proportion of correct answers among all generated answers (higher values are better).
    \item \textbf{Hallucination Rate:\cite{capellini2024knowledge}} The proportion of hallucinated answers among all generated answers (lower values are better).
    \item \textbf{Rejection Rate:\cite{chen2024benchmarking, yu2024evaluation}} The proportion of cases where the retrieved context was insufficient to answer the query (lower values are better).
    \item \textbf{Adjusted Accuracy:\cite{es2023ragas, min2022rethinkingroledemonstrationsmakes,kryscinski2019evaluating}} The proportion of correct predictions among all cases where the model made a prediction, excluding cases with insufficient context. It ensures that the metric focuses only on cases where the model attempts to answer, providing a more precise evaluation of its performance.
    This metric is defined as:
    
    \begin{equation} 
    \text{Adjusted Accuracy} = \frac{\text{Correct Answers}}{\text{Correct Answers} + \text{Hallucinated Answers}} \times 100
    \end{equation}

\end{itemize}

\subsection{Results}

\begin{figure}[!t] 
    \centering
    \includegraphics[width=0.8\textwidth]{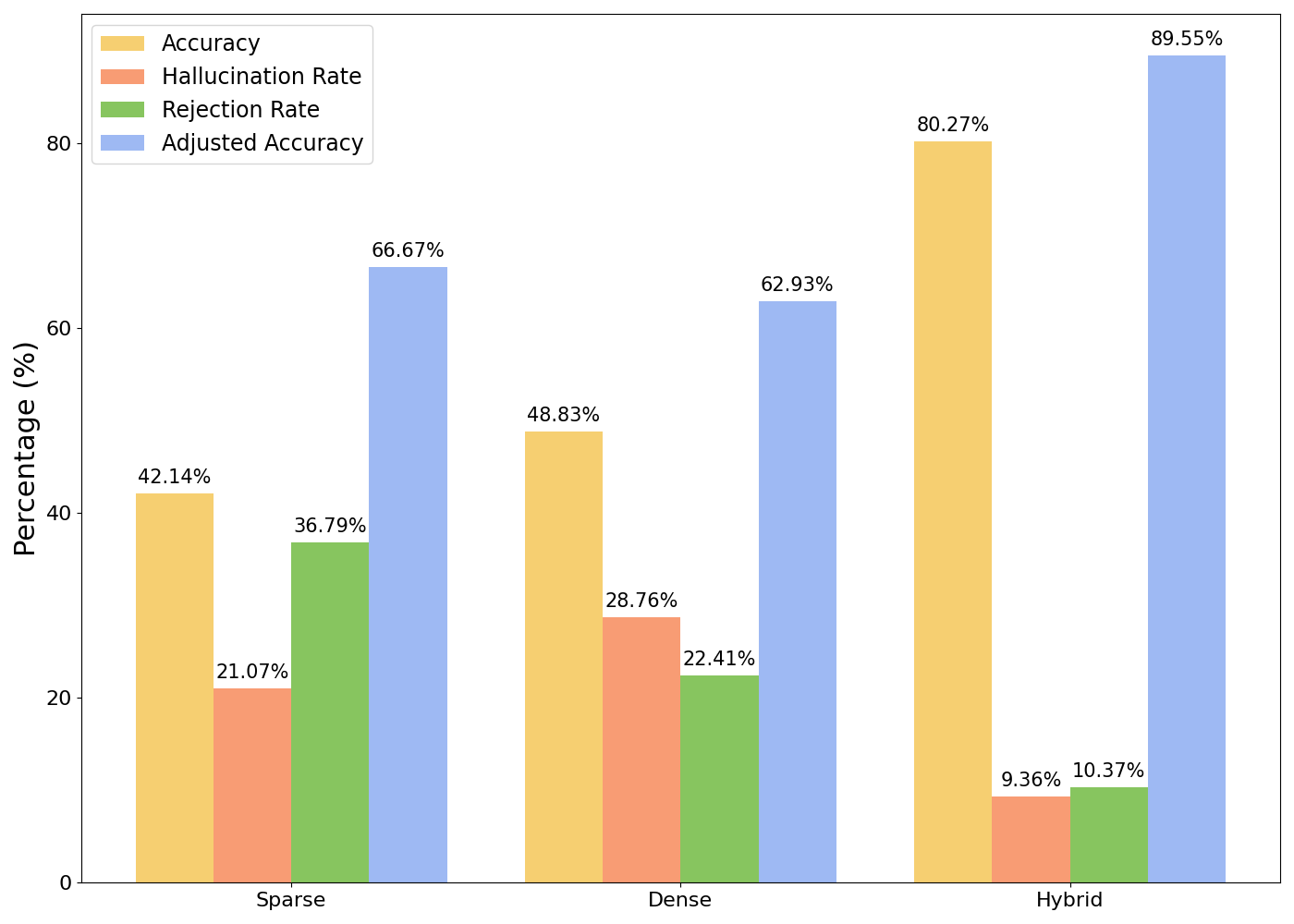} 
    \caption{Overall Performance in Mitigating Hallucinations on $HaluBench_{small}$}
    \label{fig:overall_halubench} % Label for referencing the image
\end{figure}

\paragraph{Retrieval Performance}
\label{retrieval_performance}

The evaluation results on $HaluBench_{orig}$ of the three retrievers, sparse, dense, and hybrid based on two metrics MAP@3 and NDCG@3 are displayed in Table~\ref{tab:retrieval_performance}. Regarding the MAP metric, $Ret_S$ gives a score of $0.724$, $Ret_D$ has $0.768$, while $Ret_{Hyb-{RRF}}$ achieves $0.897$. Similarly, for the NDCG, $Ret_S$ and $Ret_D$ get $0.732$ and $0.783$ respectively, whereas $Ret_{Hyb-{RRF}}$ has a relatively higher score of $0.915$. The results indicate that hybrid retriever outperforms both the sparse and dense retrievers across both retrieval metrics, demonstrating the effectiveness of combining lexical and semantic retrieval techniques. The performance gap between the retrievers in terms of NDCG is larger due to its sensitivity to ranking.
The enhancements in NDCG and MAP can be attributed to the hybrid retriever's capability to capture both exact matches and semantic relevance, along with its utilization of query expansion and dynamic weighting.

%for the sparse retriever is 0.732, for the dense retriever is 0.783, and for the hybrid retriever is \textbf{0.915} respectively. These results demonstrate that the hybrid retriever consistently outperformed the other two methods, highlighting the effectiveness of combining lexical and semantic retrieval techniques. The improvements in NDCG and MAP can be attributed to the hybrid retriever's ability to capture both exact matches and semantic relevance, as well as its use of query expansion and dynamic weighting.

\begin{table}[!h]
\centering
\begin{tabular}{lccc}
\toprule
\textbf{Metric} & \textbf{Sparse ($Ret_S$)} & \textbf{Dense ($Ret_D$)} & \textbf{Hybrid ($Ret_{Hyb-{RRF}}$)} \\
\midrule
MAP@3  & 0.724 & 0.768 & \textbf{0.897} \\
NDCG@3 & 0.732 & 0.783 & \textbf{0.915} \\
\bottomrule
\end{tabular}
\caption{Retrieval Performance Evaluation on $HaluBench_{orig}$}
\label{tab:retrieval_performance}
\end{table}

\paragraph{Overall Performance on Hallucinations}
\label{overall_performance}

To assess the overall performance of the RAG pipeline in mitigating hallucinations, we use the metrics defined in Section~\ref{overall_metrics}. This involves a comparative evaluation of three RAG pipelines with different retrieval approaches ($Ret_S$, $Ret_D$, and $Ret_{Hyb-{RRF}}$), allowing us to examine the performance of different retrieval methods in mitigating hallucinations.
In other words, this evaluation provides insights into whether they provide relevant and sufficient context for the next step in the RAG pipeline, the generation phase (Section~\ref{subsec:generation}), which aims to generate accurate responses by prompting the model $M$. The overall evaluation results on $HaluBench_{small}$ are displayed in Figure~\ref{fig:overall_halubench}.

\begin{table}[!t]
\centering
\caption{Overall Performance in Mitigating Hallucinations Across Six Source Datasets}
\begin{adjustbox}{width=\textwidth}
\begin{tabular}{@{}llcccc@{}}
\toprule
\textbf{Datasets}       & \textbf{Retrievers} & \textbf{Accuracy (\%)} & \textbf{Hallucination Rate (\%)} & \textbf{Rejection Rate (\%)} & \textbf{Adjusted Accuracy (\%)} \\ \midrule
\multirow{3}{*}{\textbf{HaluEval}} 
                        & $Ret_S$       & 56.00               & 22.00                         & 22.00                               & 71.79                        \\
                        & $Ret_D$            & 64.00               & 22.00                         & 14.00                               & 74.42                        \\
                        & $Ret_{Hyb-{RRF}}$            & \textbf{\underline{92.00}}               & \textbf{6.00}                        & \textbf{\underline{2.00}}                               & \textbf{93.88}                        \\ \midrule
\multirow{3}{*}{\textbf{Drop}}   
                        & $Ret_S$        & 30.61               & 48.98                         & 20.41                               & 38.46                        \\
                        & $Ret_D$           & 48.98               & 38.78                         & 12.24                               & 55.81                        \\
                        & $Ret_{Hyb-{RRF}}$              & \textbf{77.55}               & \textbf{14.29}                         & \textbf{8.16}                               & \textbf{84.44}                        \\ \midrule
\multirow{3}{*}{\textbf{RAGTruth}} 
                        & $Ret_S$       & 68.00               & 12.00                         & 20.00                               & 85.00                        \\
                        & $Ret_D$           & 76.00               & 10.00                         & 14.00                               & 88.37                        \\
                        & $Ret_{Hyb-{RRF}}$              & \textbf{88.00}               & \textbf{\underline{4.00}}                         & \textbf{8.00}                              & \textbf{95.65}                        \\ \midrule
\multirow{3}{*}{\textbf{PubMed}} 
                        & $Ret_S$       & 60.00               & 16.00                        & 24.00                               & 78.95                        \\
                        & $Ret_D$           & 66.00               & 20.00                         & 14.00                               & 76.74                        \\
                        & $Ret_{Hyb-{RRF}}$              & \textbf{\underline{92.00}}               & \textbf{\underline{4.00}}                         & \textbf{4.00}                               & \textbf{\underline{95.83}}                        \\ \midrule
\multirow{3}{*}{\textbf{CovidQA}} 
                        & $Ret_S$      & 30.00               & 20.00                         & 50.00                               & 60.00                        \\
                        & $Ret_D$            & 14.00               & 58.02                         & 28.10                               & 19.44                        \\
                        & $Ret_{Hyb-{RRF}}$              & \textbf{70.02}              & \textbf{\textcolor{red}{22.00}}                         & \textbf{8.00}                               & \textbf{\textcolor{red}{76.09}}                        \\ \midrule
\multirow{3}{*}{\textbf{FinanceBench}} 
                        & $Ret_S$       & 8.00               & 8.02                         & 84.00                               & 50.00                        \\
                        & $Ret_D$            & 26.00               & 24.30                         & 50.00                               & 52.00                        \\
                        & $Ret_{Hyb-{RRF}}$              & \textbf{\textcolor{red}{62.90}}               & \textbf{6.00}                        & \textbf{\textcolor{red}{32.00}}                              & \textbf{91.18}                        \\ \bottomrule
\end{tabular}
\end{adjustbox}
\label{tab:overall_sixdatasets}
\end{table}

The results show that the complete RAG pipeline using $Ret_{Hyb-{RRF}}$ as the retriever outperformed the other two pipelines, which use $Ret_S$ and $Ret_D$ retrievers, across all four evaluation metrics.
Following the overall evaluation on the annotated HaluBench dataset, we also assessed the performance of the RAG pipeline with $Ret_{Hyb-{RRF}}$ separately across six source datasets from various domains.
Based on accuracy, $Ret_{Hyb-{RRF}}$ showed the best performance on the HaluEval and PubMed datasets with the accuracy score of 92.00, while the worst performance on the FinanceBench (the accuracy score of 62.90). In terms of hallucination rate (lower is better), $Ret_{Hyb-{RRF}}$ achieved the lowest score of 4.00 on the RAGTruth and PubMed datasets, whereas the highest score of 22.00 on the CovidQA. Regarding the rejection rate (lower is better), although $Ret_{Hyb-{RRF}}$ had the best performance on the HaluEval with the rejection rate of 2.00, the results on the other datasets except the FinanceBench are similar. $Ret_{Hyb-{RRF}}$ got the highest rejection rate of 32.00 on the FinanceBench. Based on adjusted accuracy, $Ret_{Hyb-{RRF}}$ achieved the highest score on the PubMed, while the lowest on the CovidQA. The findings reveal that although $Ret_{Hyb-{RRF}}$ exhibited the poorest performance on each metric for the CovidQA and FinanceBench datasets which are domain-specific challenging datasets, it significantly enhanced the results on these datasets with respect to other two retrievers.
%and the lowest hallucination rate compared to the BM25 and semantic retrievers. The hybrid retriever's ability to provide more relevant and comprehensive context significantly reduced the hallucination rate, particularly in challenging datasets such as PubMed, CovidQA, and FinanceBench. The insufficient context rate was also minimized, indicating that the hybrid retriever not only retrieved relevant documents but also ensured that the retrieved context was sufficient for answer generation.

\begin{figure}[H] 
    \centering
    \includegraphics[width=0.8\textwidth]{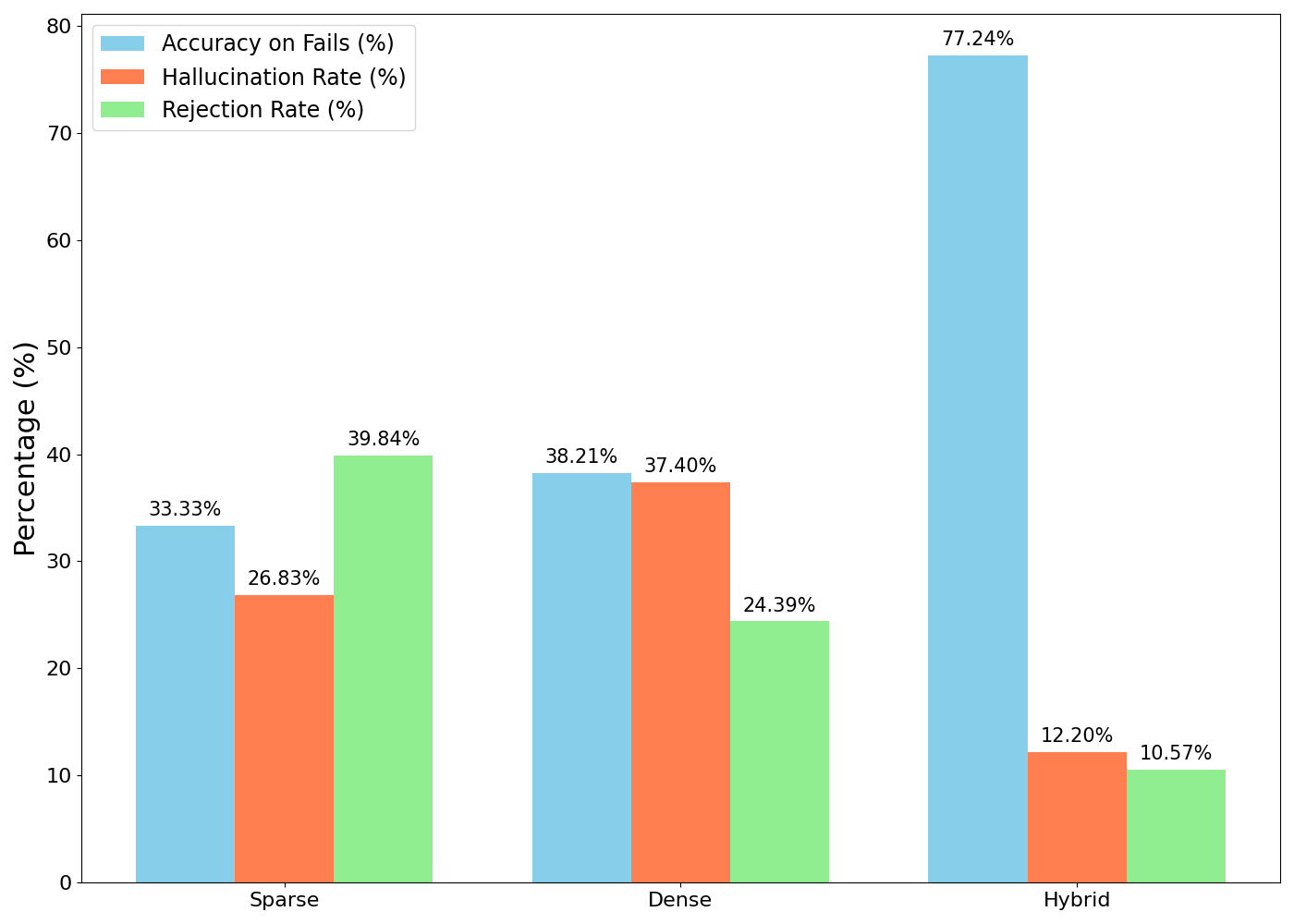} 
    \caption{Metrics comparison on only Hallucinated Samples}
    \label{fig:overall_hallucinated} % Label for referencing the image
\end{figure}

%
Then, we also evaluated the performance of the pipeline with $Ret_{Hyb-{RRF}}$ only on the hallucinated samples (labelled as $FAIL$ in the annotated dataset). There were 125 hallucinated samples in total. RAGTruth dataset does not contain any samples labelled as $FAIL$. For this, we used the same three metrics from Section~\ref{overall_metrics} as accuracy, hallucination rate and rejection rate which were computed only on the 125 hallucinated examples.
The overall evaluation results on these 125 hallucinated samples are displayed in Figure~\ref{fig:overall_hallucinated}, where the RAG pipeline with the $Ret_{Hyb-{RRF}}$ outperformed others. And the detailed evaluation results of hallucinated samples on each dataset are displayed in Appendix~\ref{hallucinated_each_dataset}. We have also evaluated our hybrid RAG pipeline by comparing it with the baseline LLM(Llama-3-instruct-8B) model, the results are illustrated in Appendix~\ref{baseline_comparision}.
The results emphasize the importance of high-quality retrieval in minimizing hallucinations in RAG systems. The hybrid retriever, combining lexical and semantic methods, provided more relevant context, improving answer generation and reducing hallucination rates.

\section{Conclusion}
In this paper we presented a hybrid retrieval approach $Ret_{Hyb-{RRF}}$, designed to mitigate hallucinations in LLMs by leveraging both sparse and dense retrievers. Experimental results on the HaluBench dataset demonstrated that hybrid retriever, which combines keyword search and semantic search methods with query expansion and dynamic weighting, consistently outperformed the other two sparse and dense retrieval methods in terms of MAP@3 and NDCG@3. Moreover, the hybrid retriever reduced hallucination rates and improved retrieval precision across domain-specific datasets, most notably in medical and financial domains which are considered as more challenging. By providing more relevant contextual documents, the hybrid strategy enabled higher accuracy in LLM-generated answers and fewer instances of insufficient context. These findings highlight the value of integrating hybrid retrieval methods for better robustness and reliability. 

Future work may further explore optimizations, including incorporating advanced re-ranking algorithms to further refine the selection of retrieved documents and by adapting our method to various data sets which are domain-specific. We also investigate the impact of the proposed method on other types of LLMs, to evaluate its broader applicability and effectiveness.

\newpage

\bibliographystyle{vancouver}
\bibliography{references}

\appendix
\section*{Appendix}
\renewcommand{\thesubsection}{\Alph{subsection}}

\section{Prompt used for the experiment}
\label{sec:prompt-details}

\texttt{[INST] You are a precise and helpful assistant. When responding:\\
- Provide a single, clear answer without repetition \\
- Don't restate the question or context. DO NOT REPEAT THE PROMPT IN THE RESPONSE AND DO NOT WRITE ANY CODE.\\
- Search if you can find the relevant answer in the provided context.\\
- If uncertain, say "The context doesn't provide sufficient information to answer the question" \\
- Avoid unnecessary formatting tokens in the response \\
- Be direct and concise while maintaining a friendly tone, avoid long explanations \\
- Only provide the answer to the question \\
Context: \{context\} \\
Question: \{question\} \\
Answer: [/INST]}

\section{Baseline comparision}
\label{baseline_comparision}
\begin{figure}[H] 
    \centering
    \includegraphics[width=0.6\textwidth]{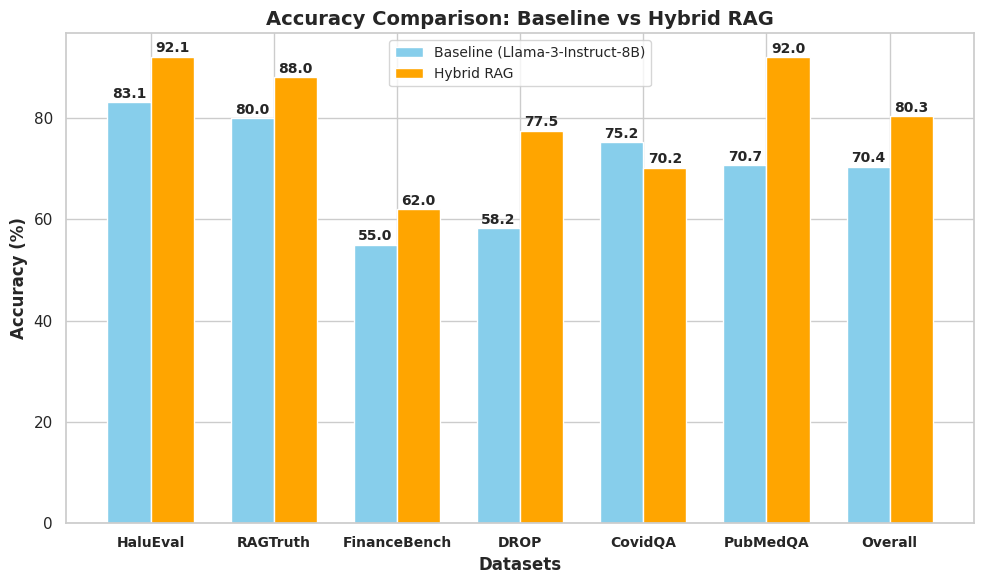} % Replace 'example-image' with your image file name
    \caption{This is an example caption for the image.}
    \label{fig:example-image} % Label for referencing the image
\end{figure}

\section{RAG metrics comparision of various datasets}

\begin{figure}[htbp]
    \centering
    % First row of images
    \begin{subfigure}[b]{0.47\textwidth}
        \includegraphics[width=\textwidth]{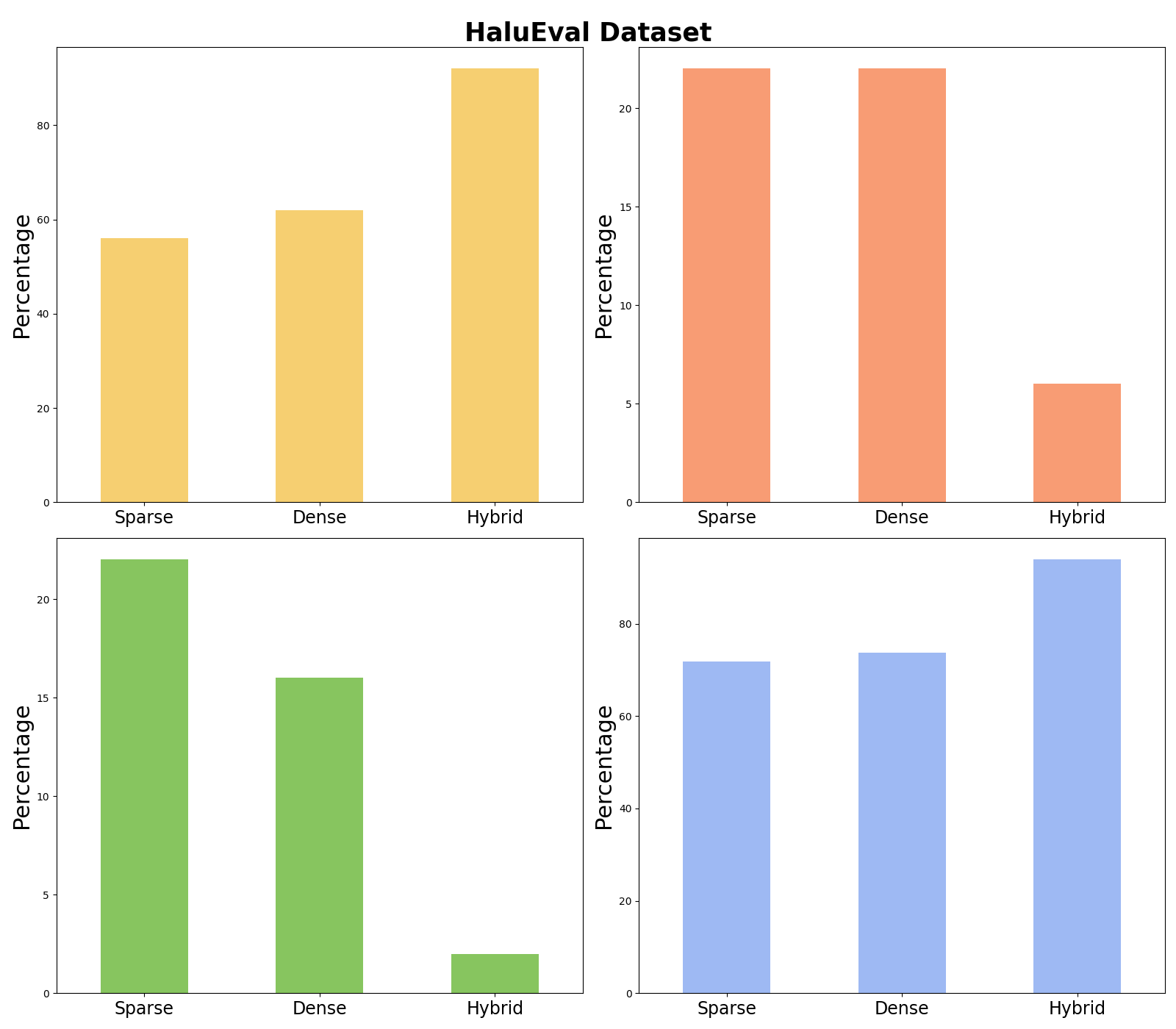}
        \caption{RAG Metrics comparision on HaluEval Dataset}
        \label{fig:image1}
    \end{subfigure}
    \hfill
    \begin{subfigure}[b]{0.47\textwidth}
        \includegraphics[width=\textwidth]{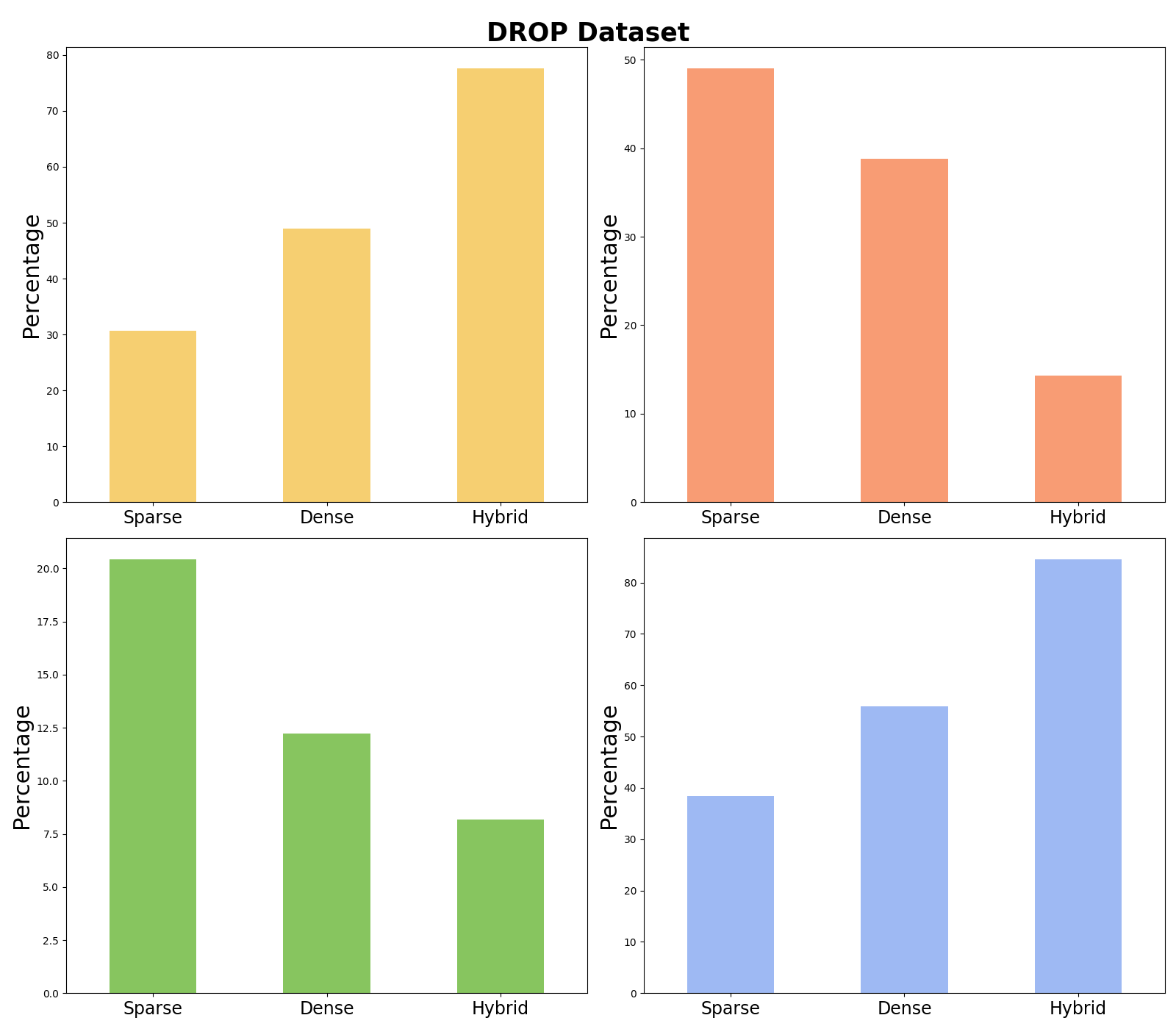}
        \caption{RAG Metrics comparision on Drop Dataset}
        \label{fig:image2}
    \end{subfigure}
    \hfill

    \vspace{1em} % Add vertical space between rows

    % Second row of images
    \begin{subfigure}[b]{0.47\textwidth}
        \includegraphics[width=\textwidth]{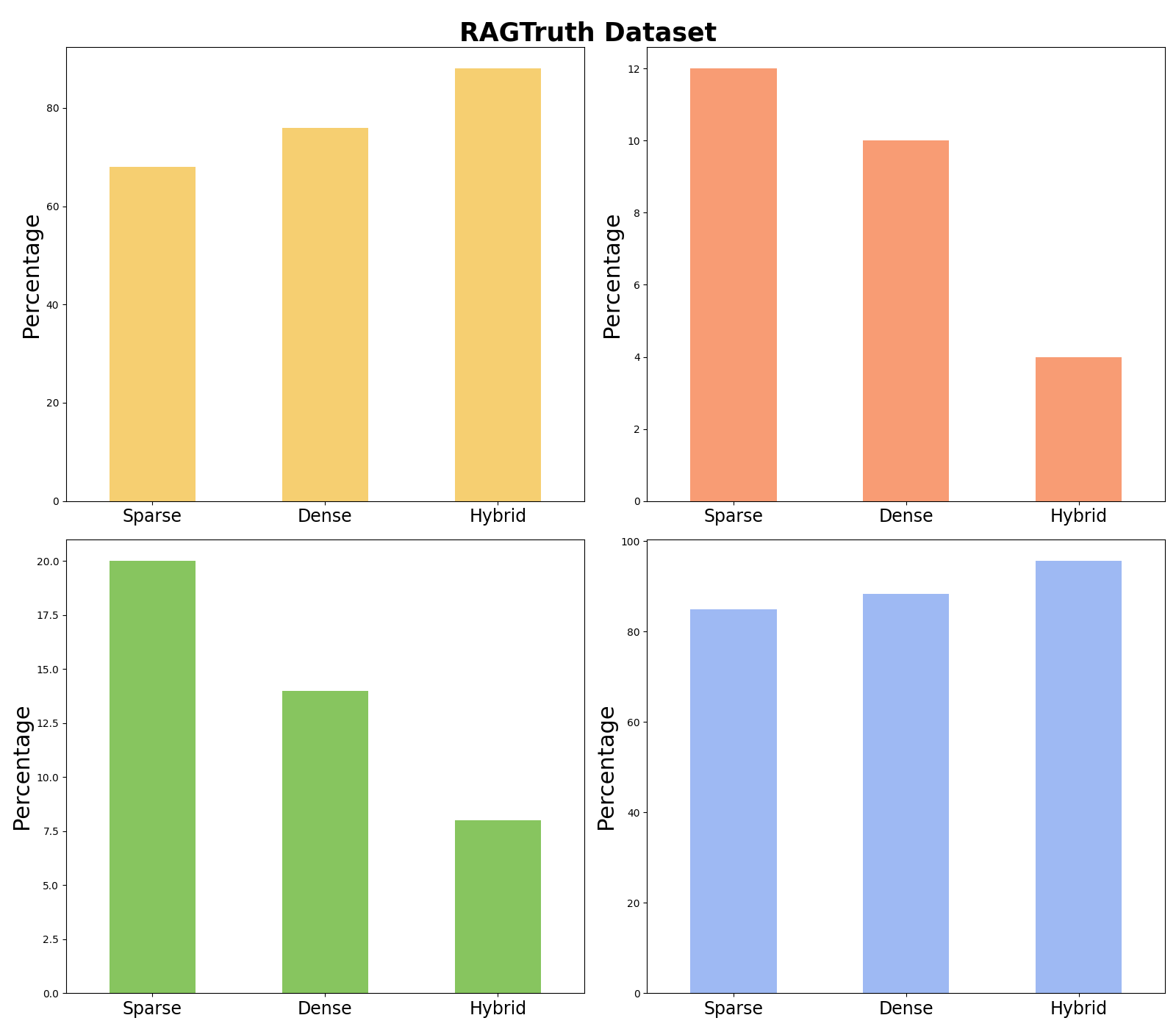}
        \caption{RAG Metrics comparision on RAGTruth Dataset}
        \label{fig:image3}
    \end{subfigure}
    \hfill    
    \begin{subfigure}[b]{0.47\textwidth}
        \includegraphics[width=\textwidth]{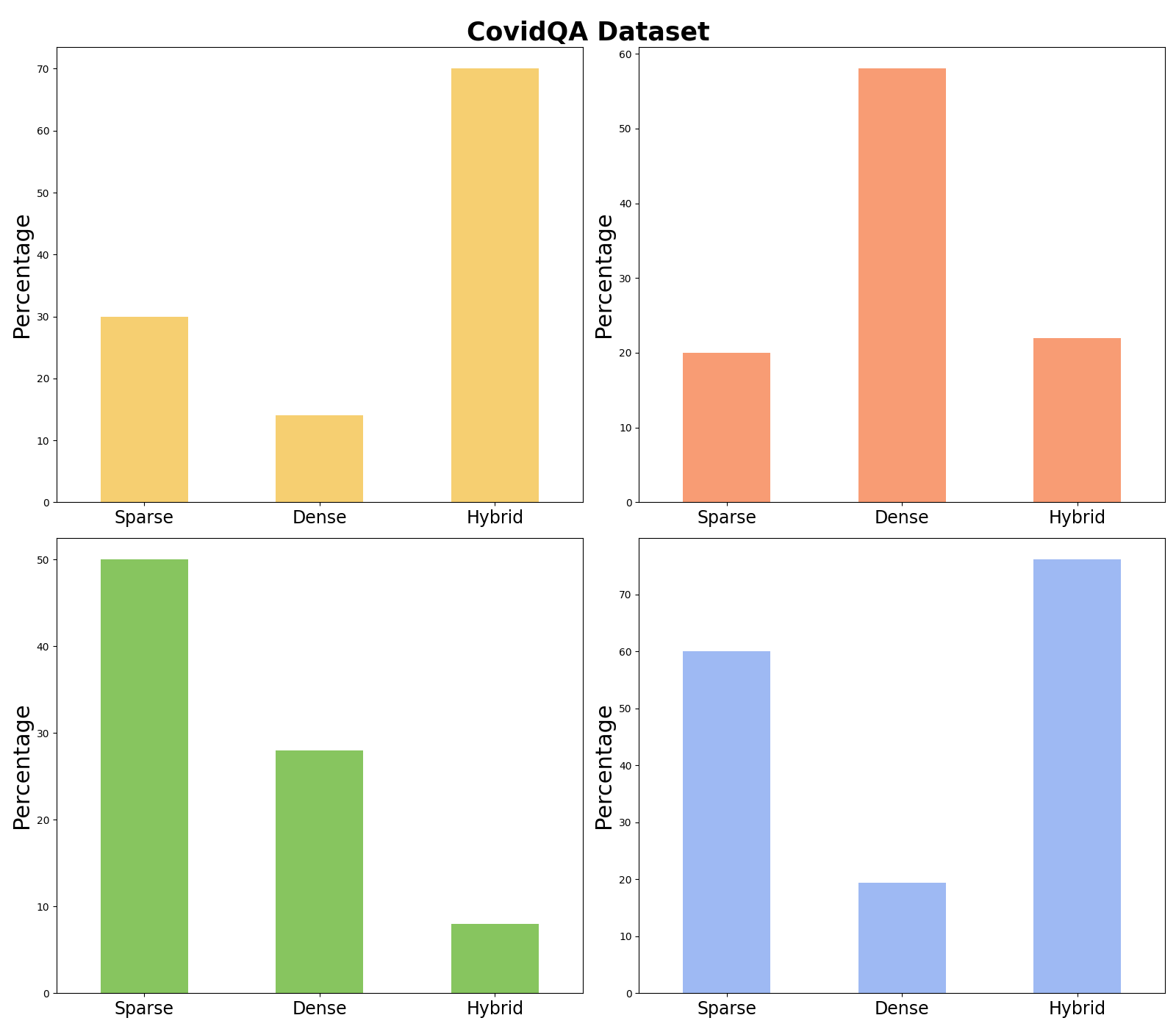}
        \caption{RAG Metrics comparision on CovidQA Dataset}
        \label{fig:image4}
    \end{subfigure}
    \hfill
    
    \vspace{1em} % Add vertical space between rows

        % Third row of images
    \begin{subfigure}[b]{0.47\textwidth}
        \includegraphics[width=\textwidth]{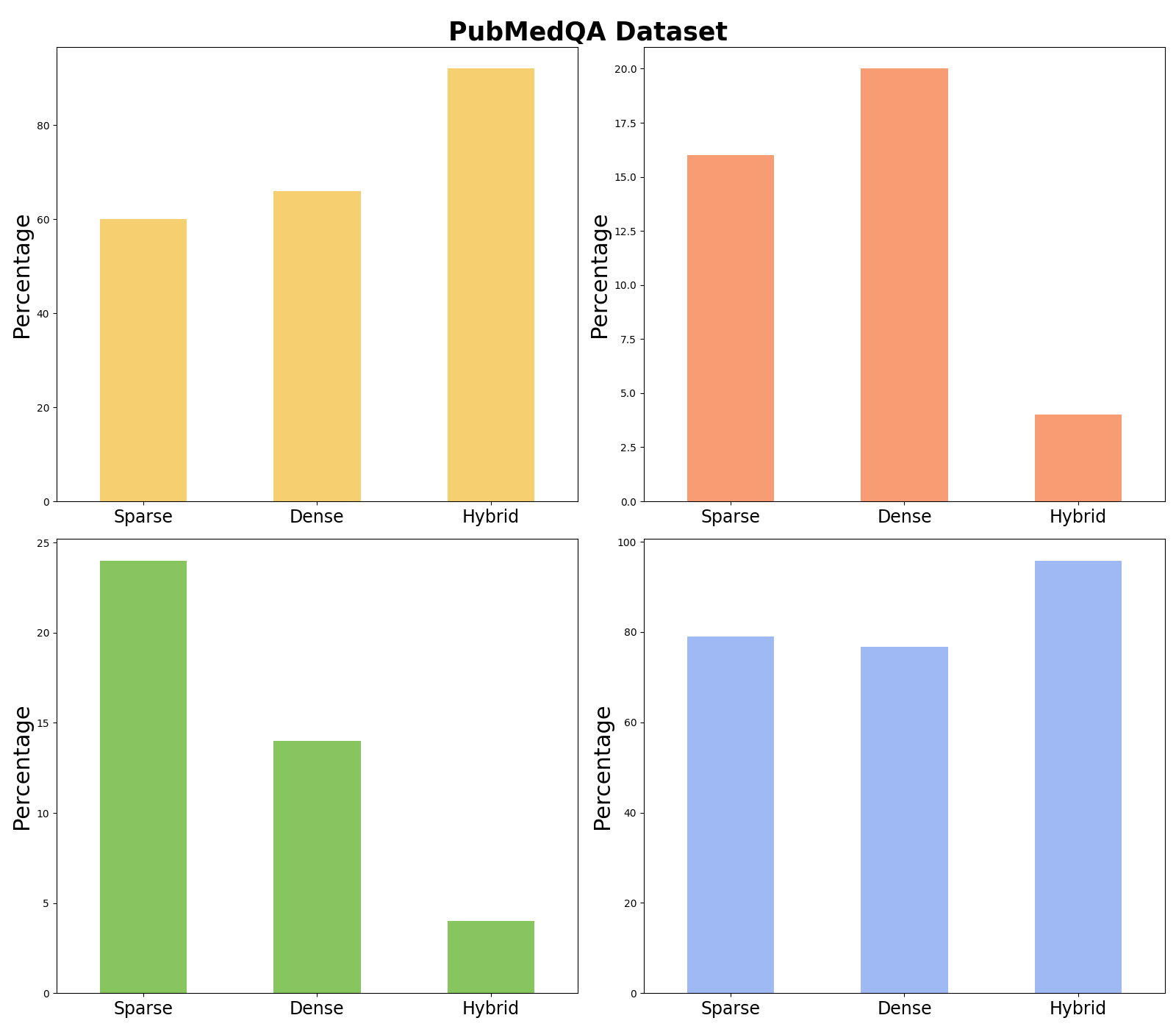}
        \caption{RAG Metrics comparision on PubMedQA Dataset}
        \label{fig:image5}
    \end{subfigure}
    \hfill
    \begin{subfigure}[b]{0.47\textwidth}
        \includegraphics[width=\textwidth]{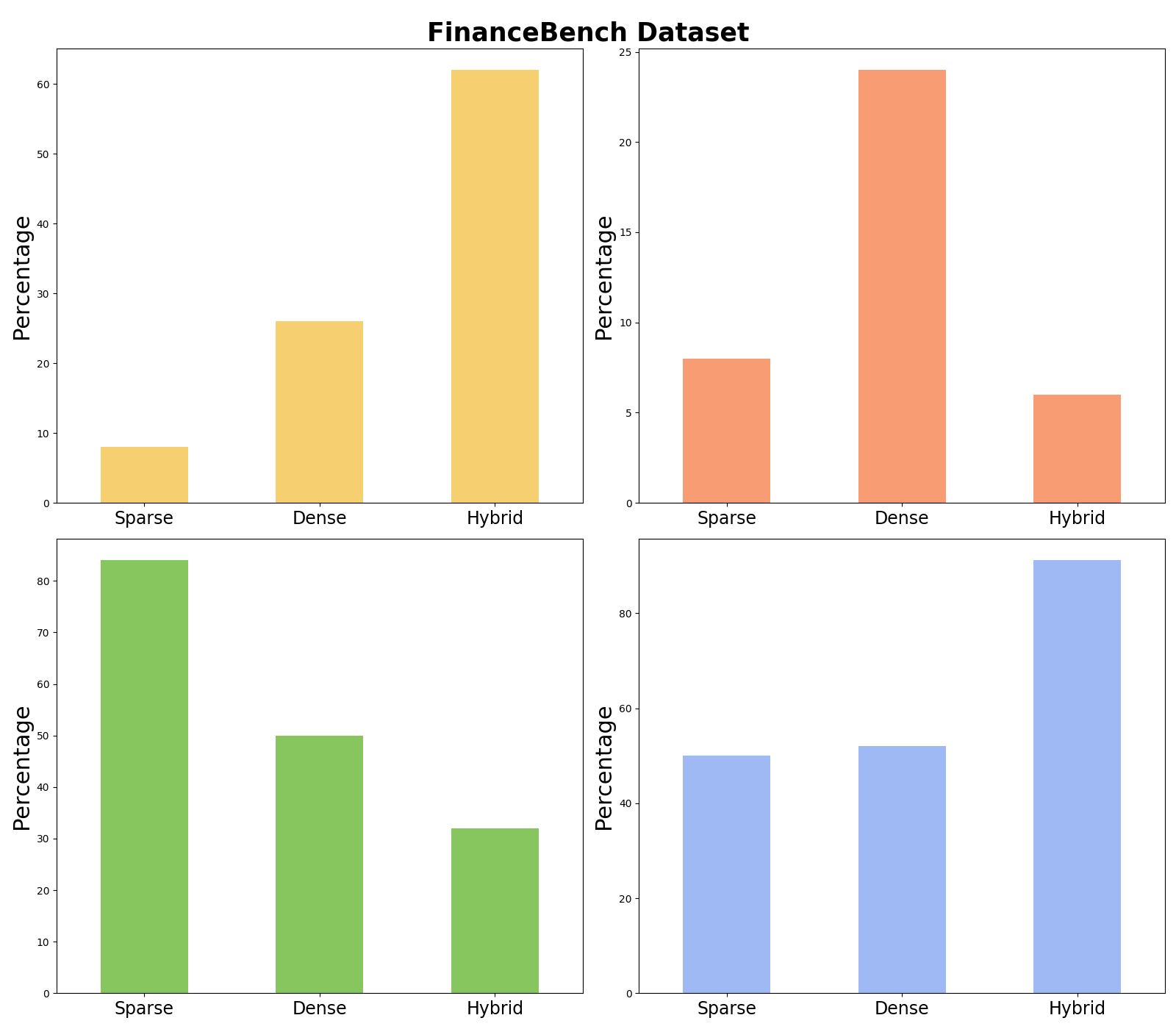}
        \caption{RAG Metrics comparision on FinanceBench Dataset}
        \label{fig:image6}
    \end{subfigure}

    \caption{RAG Metrics comparision on each dataset}
    \label{fig:collage}
\end{figure}

\section{Hallucinated Samples analysis}
\label{hallucinated_each_dataset}
\begin{figure}[htbp]
    \centering
    % First row of images
    \begin{subfigure}[b]{0.47\textwidth}
        \includegraphics[width=\textwidth]{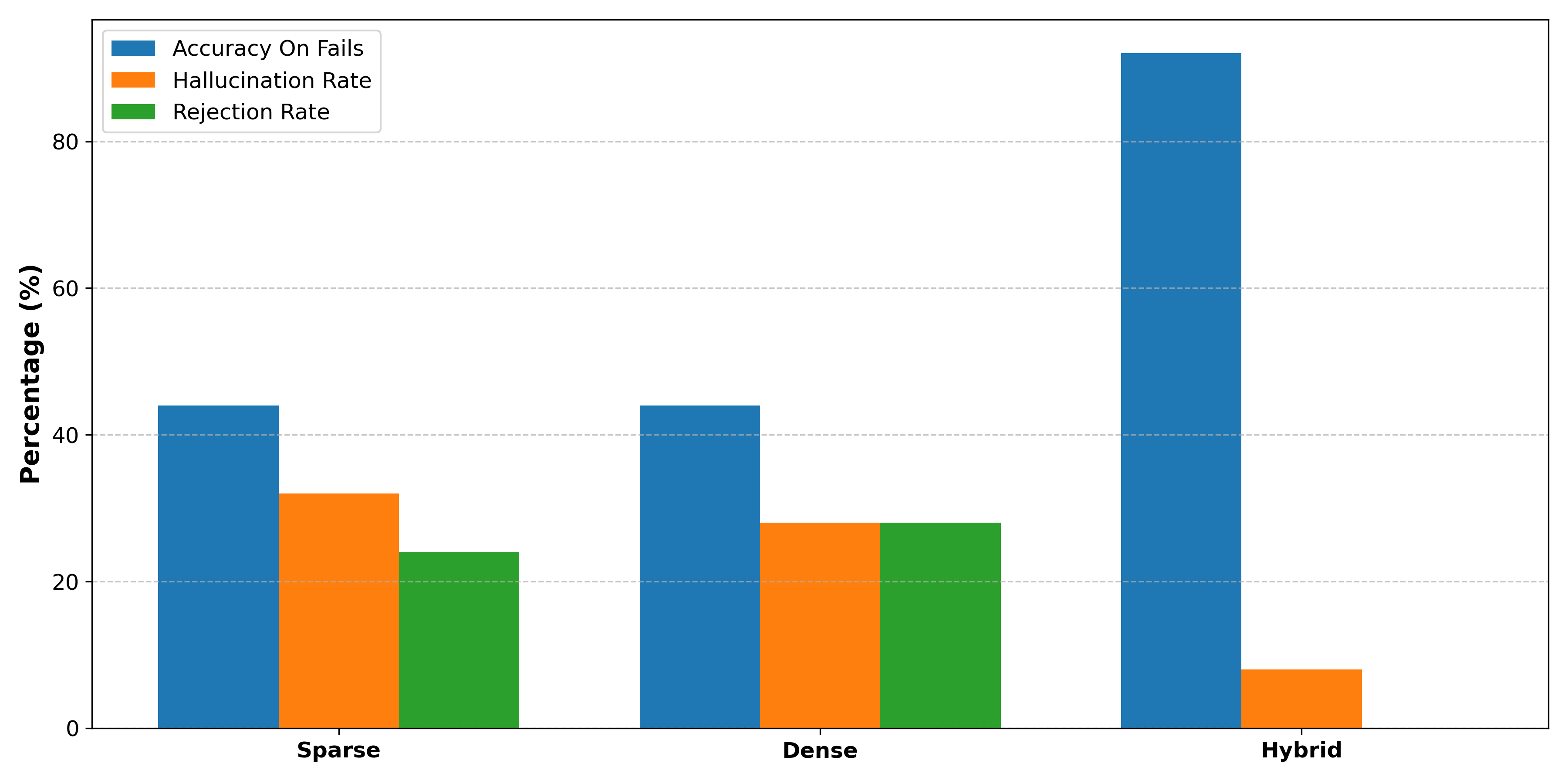}
        \caption{Hallucinated Samples analysis on HaluEval Dataset}
        \label{fig:halluimage1}
    \end{subfigure}
    \hfill
    \begin{subfigure}[b]{0.47\textwidth}
        \includegraphics[width=\textwidth]{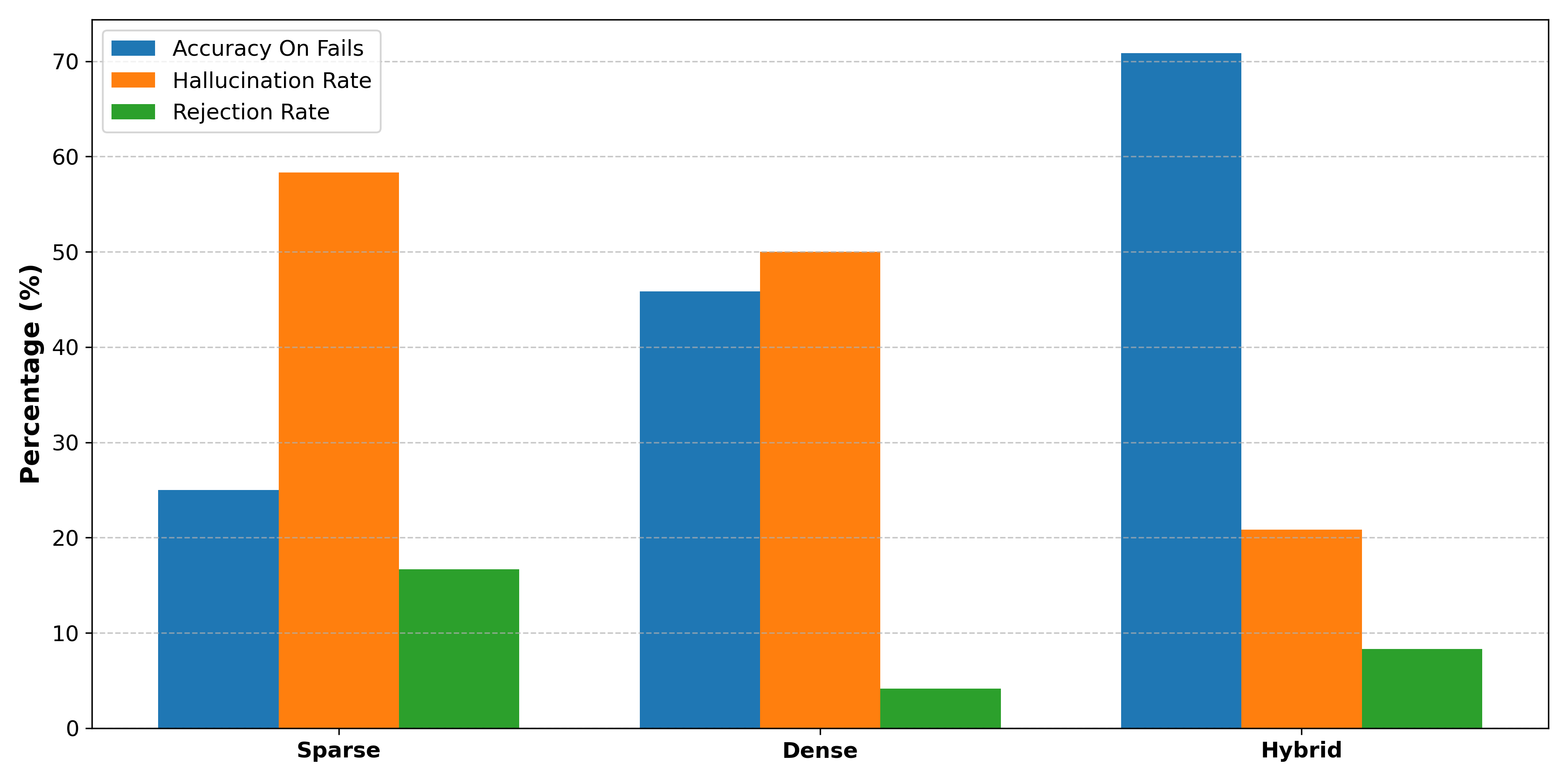}
        \caption{Hallucinated Samples analysis on Drop Dataset}
        \label{fig:halluimage2}
    \end{subfigure}
    \hfill

    \vspace{1em} % Add vertical space between rows

    % Second row of images  
    \begin{subfigure}[b]{0.47\textwidth}
        \includegraphics[width=\textwidth]{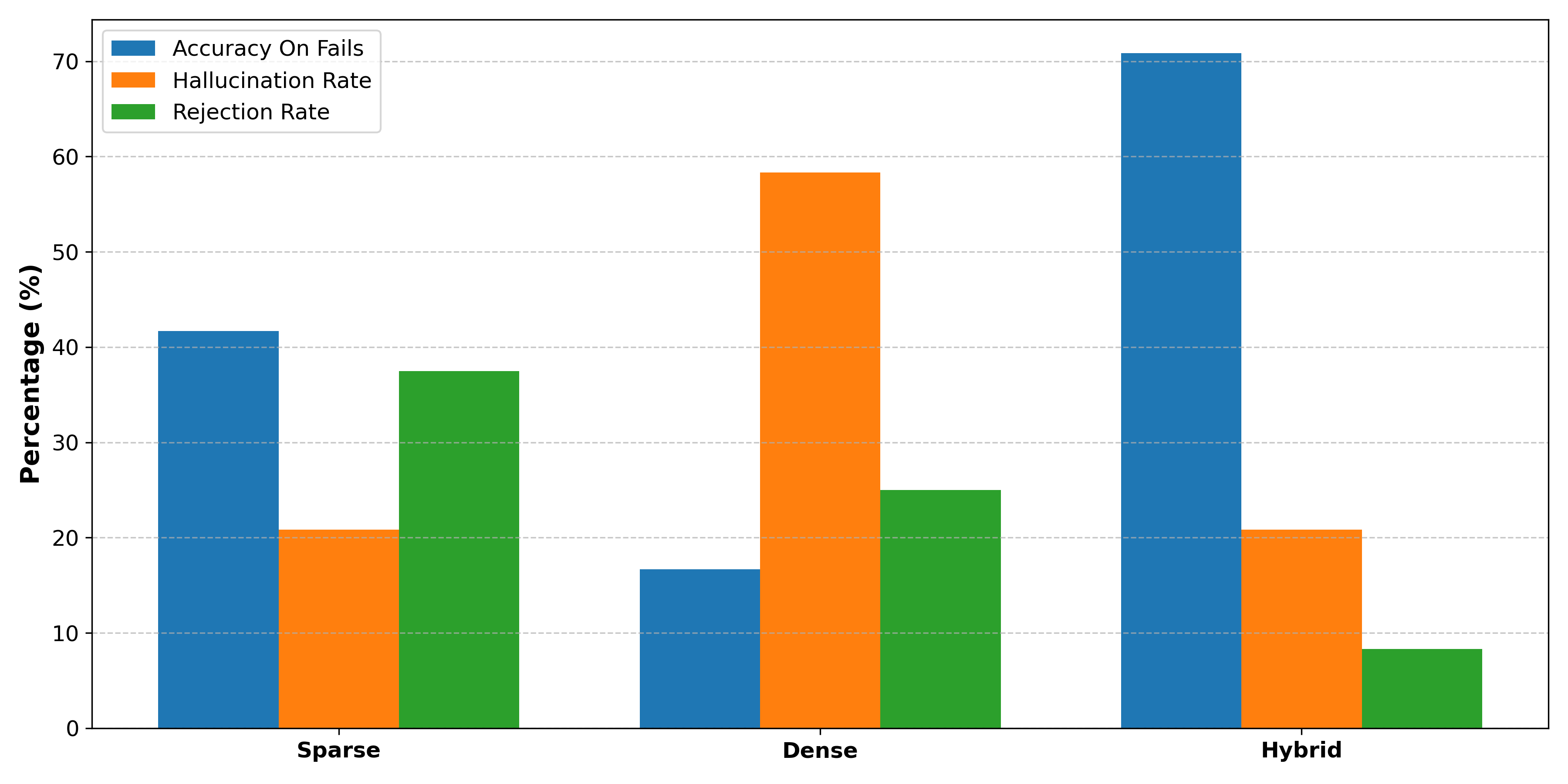}
        \caption{Hallucinated Samples analysis on CovidQA Dataset}
        \label{fig:halluimage4}
    \end{subfigure}
    \hfill
    \begin{subfigure}[b]{0.47\textwidth}
        \includegraphics[width=\textwidth]{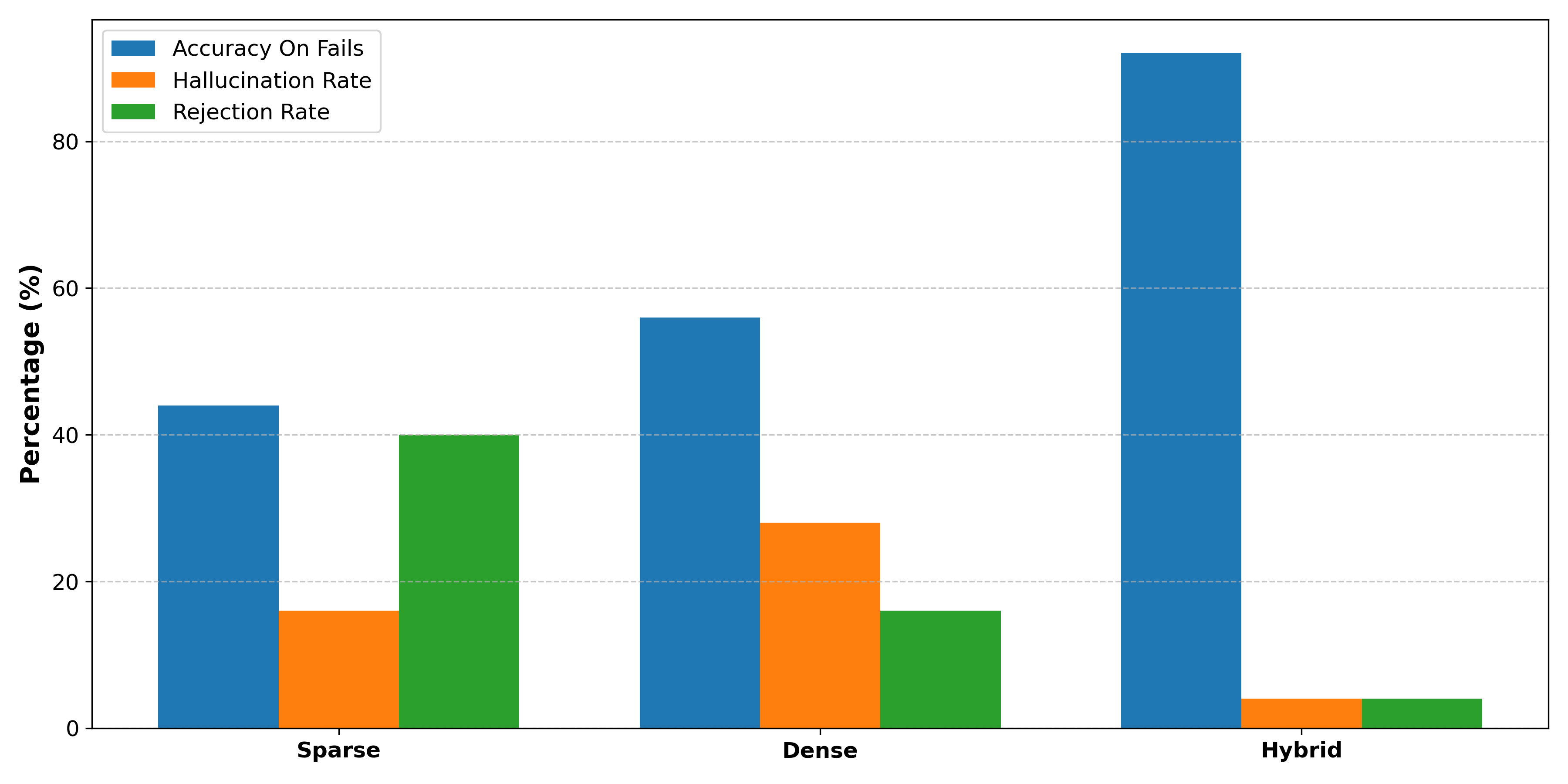}
        \caption{Hallucinated Samples analysis on PubMedQA Dataset}
        \label{fig:halluimage5}
    \end{subfigure}
    \hfill
    
    \vspace{1em} % Add vertical space between rows

        % Third row of images
    \begin{subfigure}[b]{0.47\textwidth}
        \includegraphics[width=\textwidth]{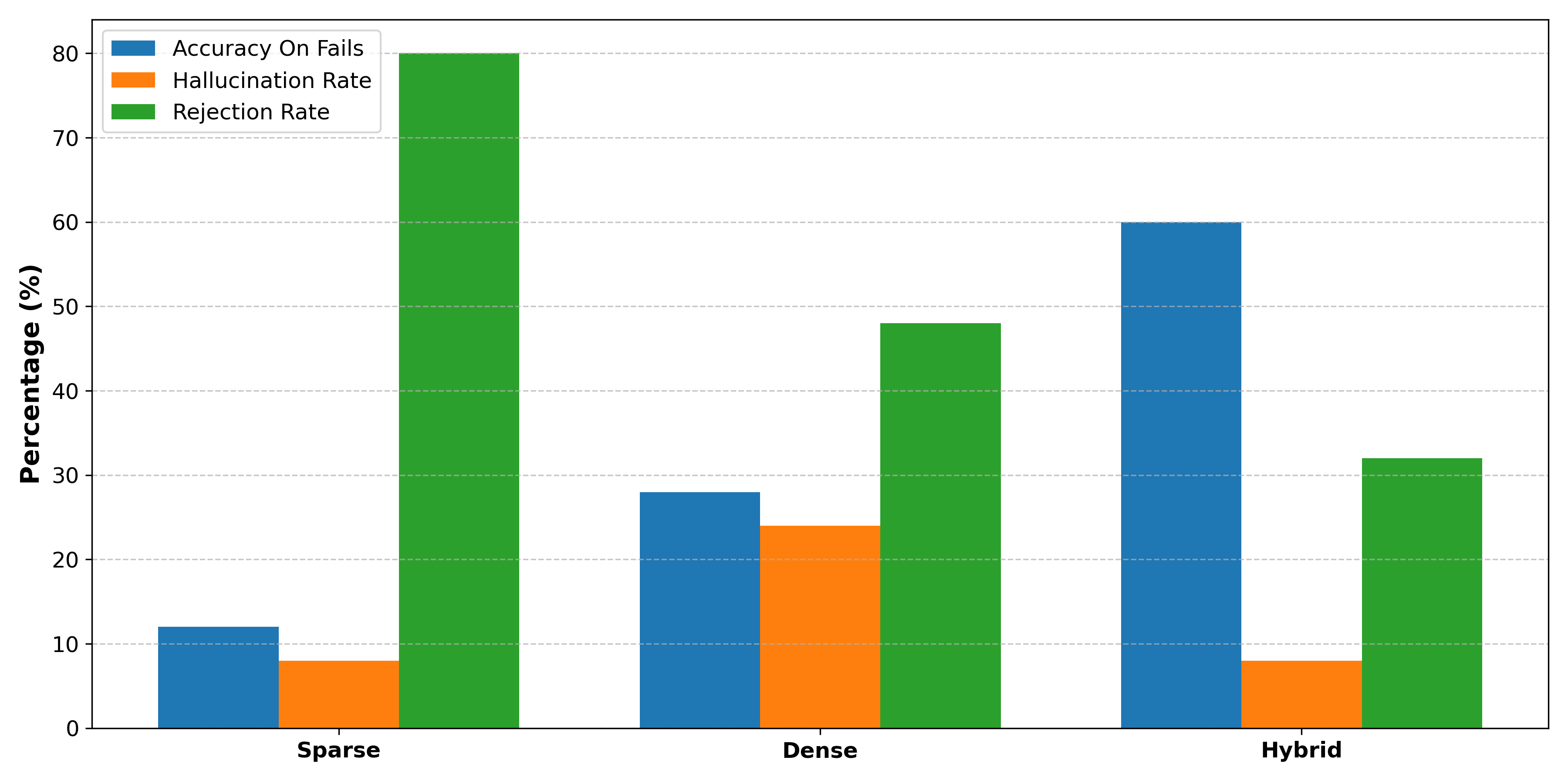}
        \caption{Hallucinated Samples analysis on FinanceBench Dataset}
        \label{fig:halluimage6}
    \end{subfigure}

    \caption{Hallucinated Samples analysis on each dataset}
    \label{fig:hallucollage}
\end{figure}

\section{Examples of hallucinations from HaluBench.}

\begin{table}[ht]
\centering
\begin{tabular}{|p{2.5cm}|p{10cm}|}
\hline
\textbf{Dataset} & \textbf{Example} \\
\hline
HaluEval & \textbf{Context}: 750 Seventh Avenue is a 615 ft (187m) tall Class-A office skyscraper in New York City. 101 Park Avenue is a 629 ft tall skyscraper in New York City, New York. \\
 & \textbf{Question}: 750 7th Avenue and 101 Park Avenue, are located in which city? \\
 & \textbf{Answer}: 750 7th Avenue and 101 Park Avenue are located in Albany, New York. \\
\hline
DROP & \textbf{Context}: Hoping to rebound from the road loss to the Chargers, the Rams went home for Week 9, as they fought the Kansas City Chiefs in a Show Me State Showdown: The Chiefs struck first as RB Larry Johnson got a 1-yard TD run for the only score of the period. In the second quarter, things got worse for the Rams as QB Damon Huard completed a 3-yard TD pass to TE Tony Gonzalez, while kicker Lawrence Tynes nailed a 42-yard field goal. St. Louis got on the board with RB Steven Jackson getting a 2-yard TD run, yet Huard and Gonzalez hooked up with each other again on a 25-yard TD strike. Rams kicker Jeff Wilkins made a 41-yard field goal to end the half. In the third quarter, QB Marc Bulger completed a 2-yard TD pass to WR Kevin Curtis for the only score of the period, yet the only score of the fourth quarter came from Huard completing an 11-yard TD pass to TE Kris Wilson. With the loss, the Rams fell to 4-4. \\
 & \textbf{Question}: Which team scored the longest field goal kick of the game? \\
 & \textbf{Answer}: Rams \\
\hline
CovidQA & \textbf{Context}: ......An important part of CDC’s role during a public health emergency is to develop a test for the pathogen and equip state and local public health labs with testing capacity. CDC developed an rRT-PCR test to diagnose COVID-19. As of the evening of March 17, 89 state and local public health labs in 50 states...... \\
 & \textbf{Question}: What kind of test can diagnose COVID-19? \\
 & \textbf{Answer}: rRT-PCR test \\
\hline
FinanceBench & \textbf{Context}: Consolidated Statement of Income PepsiCo, Inc. and Subsidiaries Fiscal years ended December 29, 2018, December 30, 2017 and December 31, 2016 (in millions except per share amounts) 2018 2017 2016 Net Revenue \$ 64,661...... \\
 & \textbf{Question}: What is the FY2018 fixed asset turnover ratio for PepsiCo? Fixed asset turnover ratio is defined as: FY2018 revenue / (average PP\&E between FY2017 and FY2018). Round your answer to two decimal places. \\
 & \textbf{Answer}: 3.7\% \\
\hline
PubmedQA & \textbf{Context}: ......The study cohort consisted of 1,797 subjects (1,091 whites and 706 blacks; age = 21-48 years) enrolled in the Bogalusa Heart Study since childhood. BP variability was depicted as s.d. of 4-8 serial measurements in childhood...... \\
 & \textbf{Question}: Is adult hypertension associated with blood pressure variability in childhood in blacks and whites : the bogalusa heart study? \\
 & \textbf{Answer}: No. Increases in BP variations as well as levels in early life are not predictive of adult hypertension, which suggests that childhood BP variability does not have a significant impact on the natural history of essential hypertension. \\
\hline
\end{tabular}
\caption{Different Examples from HaluBench Dataset}
\label{tab:dataset_examples}
\end{table}

\end{document}